\documentclass[pre,superscriptaddress,tightenlines,onecolumn,showpacs,longbibliography,preprint,floatfix,nofootinbib]{revtex4-1}

\pdfoutput=1

\usepackage{latexsym,amsmath,amsfonts,tabularx,amssymb,bm,empheq}
\usepackage{subfigure}

\usepackage{color}
\usepackage[dvipsnames]{xcolor}
\definecolor{nblue}{RGB}{28,130,185}
\definecolor{cgreen}{RGB}{76,153,0}
\definecolor{myorange}{RGB}{245,156,74}

\usepackage{hyperref}
\hypersetup{
  colorlinks=true,
  citecolor=magenta,
  urlcolor=-myorange
}

\newcommand{\bea}{\begin{eqnarray}}
\newcommand{\eea}{\end{eqnarray}}

\newcommand{\ddroit}[2]{\frac{\mathrm{d}#1}{\mathrm{d}#2}}

\newcommand{\ddt}[1]{\mathrm{d}#1/\mathrm{d}t}
\newcommand{\ddthat}[1]{\mathrm{d}#1/\mathrm{d}\hat t}

\newcommand{\corot}{\frac{\mathrm{D}}{\mathrm{D}t}}

\newcommand{\dd}{\mathrm{d}}

\definecolor{nblue}{RGB}{28,130,185}

\definecolor{cgreen}{RGB}{76,153,0}

\definecolor{myorange}{RGB}{245,156,74}

\def\simge{\mathrel{%
   \rlap{\raise 0.511ex \hbox{$>$}}{\lower 0.511ex \hbox{$\sim$}}}}
\def\simle{\mathrel{
   \rlap{\raise 0.511ex \hbox{$<$}}{\lower 0.511ex \hbox{$\sim$}}}}

\def\simle{\mathrel{
   \rlap{\raise 0.511ex \hbox{$<$}}{\lower 0.511ex \hbox{$\sim$}}}}

\def\simge{\mathrel{%
    \rlap{\raise 0.511ex \hbox{$>$}}{\lower 0.511ex \hbox{$\sim$}}}}

\usepackage{ulem}
\usepackage{tensor}

\usepackage{blkarray}
\usepackage{mathtools}
\usepackage{multirow}

\makeatletter
\newcommand\footnoteref[1]{\protected@xdef\@thefnmark{\ref{#1}}\@footnotemark}
\makeatother

\AtBeginDocument{%
    \newwrite\bibnotes
    \def\bibnotesext{Notes.bib}
    \immediate\openout\bibnotes=\jobname\bibnotesext
    \immediate\write\bibnotes{@CONTROL{REVTEX41Control}}
    \immediate\write\bibnotes{@CONTROL{%
    apsrev41Control,author="08",editor="1",pages="1",title="0",year="1"}}
     \if@filesw
     \immediate\write\@auxout{\string\citation{apsrev41Control}}%
    \fi
}%

\newcommand{\titleMacro}{Hydraulic and electric control of cell spheroids}

\begin{document}

\title{\titleMacro}

\author{Charlie Duclut}\email{duclut@pks.mpg.de}
\affiliation{Max-Planck-Institut f\"ur Physik Komplexer Systeme, N\"othnitzer Str. 38, 01187 Dresden, Germany.}

\author{Jacques Prost}\email{Jacques.Prost@curie.fr}
\affiliation{Laboratoire Physico Chimie Curie, UMR 168, Institut Curie, PSL Research University, CNRS, Sorbonne Universit\'e, 75005 Paris, France.}\affiliation{Mechanobiology Institute, National University of Singapore, 117411 Singapore.}

\author{Frank J\"ulicher}\email{julicher@pks.mpg.de}
\affiliation{Max-Planck-Institut f\"ur Physik Komplexer Systeme, N\"othnitzer Str. 38, 01187 Dresden, Germany.}%
\affiliation{Center for Systems Biology Dresden, Pfotenhauerstr. 108, 01307 Dresden, Germany.}%
\affiliation{Cluster of Excellence Physics of Life, TU Dresden, 01062 Dresden, Germany.}

\date{\today}

\begin{abstract}
We use a theoretical approach to examine the effect of a radial fluid flow or electric current on the growth and homeostasis of a cell spheroid. Such conditions may be generated by a drain of micrometric diameter. To perform this analysis, we describe the tissue as a continuum. We include active mechanical, electric, and hydraulic components in the tissue material properties. We consider a spherical geometry and study the effect of the drain on the dynamics of the cell aggregate. We show that a steady fluid flow or electric current imposed by the drain could be able to significantly change the spheroid long-time state. In particular, our work suggests that a growing spheroid can systematically be driven to a shrinking state if an appropriate external field is applied. Order-of-magnitude estimates suggest that such fields are of the order of the indigenous ones. Similarities and differences with the case of tumors and embryo development are briefly discussed.
\end{abstract}

\maketitle

\section*{Introduction}

Understanding how cells collectively organize to form complex structures and organs is the fundamental question raised in morphogenesis. This self-organization stems from the interplay of biochemical~\cite{green2015positional,miller2013interplay}, mechanical~\cite{miller2013interplay,mammoto2013mechanobiology,ladoux2017mechanobiology}, but also hydraulic~\cite{ruiz-herrero2017organ,dumortier2019hydraulic} and electrical processes~\cite{levin2018bioelectric,silver2018bioelectric}.
A long-standing paradigm in developmental biology is that cell chemical signals, in the form of morphogens, control cell growth and differentiation leading to tissue patterning~\cite{green2015positional}.
Although this biochemical signaling is of paramount importance 
in developmental control, it is now well-established that mechanical forces between cells or mediated by the extra-cellular matrix can also provide regulatory cues that are equally important~\cite{mammoto2013mechanobiology}.

The crucial importance of hydraulics in morphogenesis, which should not come as a surprise given the large water content in tissues, has also been highlighted in multiple experiments. 
Hydraulic oscillations have for instance been shown to provide a robust mechanism for size control of the mouse embryo~\cite{chan2019hydraulic} and during the \textit{Hydra} regeneration~\cite{futterer2003morphogenetic}.
The role of electrical signals in tissue patterning, although already studied by Roux~\cite{roux1892uber} at the end of the 19th century, has gained a new interest only recently~\cite{mccaig2005controlling,levin2018bioelectric,mclaughlin2018bioelectric}. In addition to its key function in receiving and relaying sensory information in the nervous system~\cite{watanabe2009cnidarians}, bioelectricity has been shown to have a dramatic importance in large-scale patterning: an alteration of the electrical signaling in \textit{Planaria} regeneration causes for instance the emergence of animals with multiple heads~\cite{levin2018bioelectric}. Similarly, it has recently been observed that an external electric field can be used to reverse the morphogenetic fate of \textit{Hydra}~\cite{braun2019electricinduced}.

In the quest of understanding how these different mechanisms come together to shape tissues and organs, simple cell aggregates such as spheroids have offered an appealing territory to observe tissue development and to formulate hypotheses on the underlying mechanisms. Remarkably, a rich behavior is observed even in single-cell type spheroids, as for instance their ability to pump fluid and form liquid-filled lumens~\cite{martin-belmonte2008cellpolarity,debnath2002role}. More recently, organoids have focused an increasing attention as they can recapitulate complex morphogenetic processes in a relatively simple and controllable environment~\cite{huch2015modeling,dahl-jensen2017physics}.

To unravel the connections between biochemical signals and tissue mechanics, mechanical perturbations of organoids and cell spheroids can be performed. For instance, atomic force microscopy has been used to probe the mechanical properties of mammary organoids~\cite{alcaraz2008laminin} and revealed the importance of both extracellular matrix stiffness and laminin signaling to maintain tissue integrity. 
Perturbation of the osmotic pressure around cell spheroids using large molecules of Dextran has also highlighted the importance of isotropic stress in tissue growth~\cite{montel2011stress,delarue2013mechanical}.

A broad understanding of tissue mechanics therefore requires to consider tissue electrohydraulic properties and to
be able to perturb tissues by electric or hydraulic means. In this theoretical work, we study the response of a cell spheroid to electric and hydraulic perturbations. We propose an experimental setup where a pipette or drain is used to impose a fluid flow or an external electric current through a micrometric drain. Our work shows that one could control the size of a cellular assembly using such a setup.

Our theoretical approach relies on generic features of the physics of the electro-hydro\-dynamic phenomena at play within tissues. We adopt a coarse-grained approach of tissues, which describes many cells and their micro-environments as a continuum with active material  properties~\cite{kruse2005generic,marchetti2013hydrodynamics,ranft2010fluidization}. Cell mechanical characteristics, fluid pumping, ion transport and electrical properties are thus considered in a unified framework~\cite{sarkar2019field,duclut2019fluid} that we use to analyze spheroid response to an external perturbation.

In particular, we highlight in the following that steady external flow or electric current imposed by a drain can systematically drive a proliferating spheroid to degeneracy. Such technique could be relevant in a medical context where it could be used to suppress cancerous tumors. 

Methods for suppressing malignant tumors are numerous: a first path to control tumor size is to use drugs to disturb the chemical regulation of cancerous cells to prevent them from proliferating (chemotherapy). Immunotherapy has offered a strong alternative in mitigating cancer by stimulating the immune system to suppress tumors~\cite{scott2012antibody}. In addition, radiotherapy -- which allows the suppression of the tumor by damaging the genetic material of cancerous cells through radiations -- has also proved effective for tumor removal~\cite{baskar2012cancer}.

Physics-based methods can also be used to provide new techniques for directly suppressing the cancerous tissues. High-intensity focused ultrasounds can for instance be used to locally overheat the cancerous cells~\cite{al-bataineh2012clinical}. Recent experiments also suggest that lower intensity ultrasound waves could be used to strain and suppress mechanically cancerous tissues~\cite{mittelstein2020selective,tijore2020ultrasoundmediated}, and low-frequency ultrasounds could be used to increase selectivity~\cite{tijore2020ultrasoundmediated}. Electrical perturbations have also been used: cancerous tissues can for instance be successfully suppressed using irreversible electroporation~\cite{miller2005cancer} by applying large voltage pulses in the tissue. More recently, micro-electrodes have been used in electrolytic ablation methods to locally change pH and kill a cancerous mass~\cite{perkons2018electrolytic}.

Compared to other treatments and ablation techniques~\cite{knavel2013tumor}, 
direct electrohydraulic perturbations of cancerous tissues have however remained largely unexplored. With the theoretical study that we present here, we aim at highlighting potential novel methods for controlling size of cell spheroids, that could also be used to suppress cancerous tumors.

\section{Continuum model of a spheroid with a drain}

\begin{figure}[t]
	\centering
	\null \hfill
	\subfigure[\label{fig_notation_in}]
    {\includegraphics[width=0.4\linewidth]{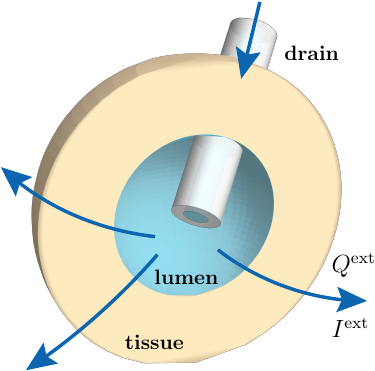}}
    \hfill
    \subfigure[\label{fig_notation_out}]
    {\includegraphics[width=0.4\linewidth]{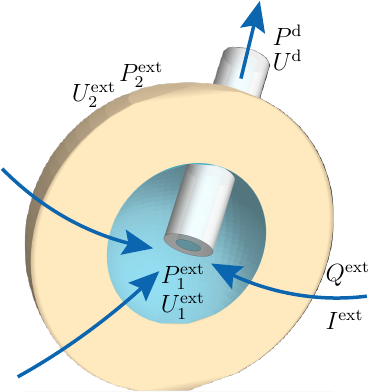}}
    \hfill \null
	\caption{Sketch of a spheroid with drain. The drain is used to impose an electric current $I^{\rm ext}$ or a volumetric flow rate $Q^{\rm ext}$, driving a fluid flow or current density throughout the cell spheroid. \textbf{(a)} Flows from inside of the spheroid to outside correspond to $Q^{\rm ext}, I^{\rm ext}>0$. \textbf{(b)} Flows from outside to inside correspond to $Q^{\rm ext}, I^{\rm ext}<0$. The pressure inside the lumen, outside the spheroid and at the outer end of the drain are denoted $P_1^{\rm ext},P_2^{\rm ext}$ and $P^{\rm d}$, respectively, and correspondingly for the electric potential: $U_1^{\rm ext}, U_2^{\rm ext}$ and $U^{\rm d}$.
	}
    \label{fig_notation}
\end{figure}

Following Refs.~\cite{sarkar2019field,duclut2019fluid}, we consider the tissue at a coarse-grained level, such that individual cells are not described but the tissue as a whole is studied as a continuum material with active electrical, hydraulic and mechanical properties. To capture the hydraulic properties of the tissue, we also adopt a two-fluid description, where the cells, which form the first fluid, are permeated by the interstitial fluid~\cite{ranft2012tissue}. In the long-time limit (several days or weeks) that we consider here, cells are able to reorganize and to relax the internal stresses within the tissue, such that an effective description of the tissue as a viscous active fluid at long times is used~\cite{ranft2010fluidization}.

To analyze specifically the effect of a drain on a cell spheroid, we consider a spherical tissue of radius $R_2$ enclosing a spherical lumen of radius $R_1$ (see Fig.~\ref{fig_notation}). A drain of inner radius $R_{\rm d}$ is inserted inside the spheroid and can be used to impose an external flow or an external current. We consider, despite the presence of the drain, a system with spherical symmetry. The description we propose in the following is therefore effectively one dimensional and depends only on the distance $r$ to the spheroid center.

Directional ion pumping through the spheroid is achieved if cells have a polarity. We thus define a cell polarity field $\bm p$ with unit norm that we assume, for simplicity, to be oriented along the radial direction: $\bm p = \bm e_r$. Cells also display a nematic ordering -- due for instance to an anisotropy in their shape -- that we describe with the nematic tensor $q_{\alpha\beta}$ (Greek indices indicate Cartesian coordinates). We assume in the following that the nematic ordering is defined by the same preferred axis as the polarity of the cells: \mbox{$q_{\alpha\beta} = p_\alpha p_\beta - (1/3)\delta_{\alpha\beta}$}. This situation is for instance observed in colon carcinoma cell spheroids~\cite{dolega2017celllike}. 
Finally, the consideration of ion pumping within the tissue requires the introduction of an electric field $\bm E$ and an electric current density $\bm j$, which obey a generalized Ohm law as we discuss in the following.

    \subsection{Tissue mechanical, hydraulic and electrical properties}

The total tissue stress is decomposed as $\sigma_{\alpha\beta} = \sigma^{\rm c}_{\alpha\beta} + \sigma^{\rm f}_{\alpha\beta}$ where here and in the following the superscripts $\rm c,f$ stand for the cells and interstitial fluid contributions, respectively. Neglecting inertia and in the absence of external bulk forces, force balance within the tissue reads:%
\begin{subequations}%
\begin{align}%
\partial_\beta\sigma^{\rm c}_{\alpha\beta} + f_\alpha &= 0 \, , \\
\partial_\beta\sigma^{\rm f}_{\alpha\beta} - f_\alpha &= 0 \, , \label{eq_forceBalance_fluid}
\end{align} \label{eq_forceBalance}%
\end{subequations}%
where $f_\alpha$ represents internal forces between the cells and the interstitial fluid, and where summation over repeated indices is implied.

To obtain the dynamics of the spheroid, that is, the time evolution of the inner and outer radii $R_1$ and $R_2$, we now need to specify the material properties (or constitutive equations) of the tissue.
In our coarse-grained description, the interstitial fluid is driven by pressure gradients, while fluid viscosity contributes to the internal forces $f_\alpha$ between fluid and cells. We therefore write the fluid stress as $\sigma^{\rm f}_{\alpha\beta}=-P^{\rm f}\delta_{\alpha\beta}$.
For the cell stress, the description of the material properties is made easier by decomposing the stress into a traceless symmetric part $\tilde\sigma^{\rm c}_{\alpha\beta}$ and an isotropic part $\sigma^{\rm c}\delta_{\alpha\beta}$. The constitutive equation for the isotropic part then reads~\cite{sarkar2019field,duclut2019fluid}:
\begin{subequations}
\begin{align}
\sigma^{\rm c} + P^{\rm c}_{\rm h} = \bar\eta
v^{\rm c}_{\gamma\gamma} - \nu_0 \tilde\sigma^{\rm c}_{\alpha\beta}q_{\alpha\beta} - \nu_1 p_\alpha E_\alpha  - \nu_2 p_\alpha V_\alpha \, , \label{eq_sigma}
\end{align}
\sloppy
where $-\sigma^{\rm c}$ is the local pressure and we have introduced the cell strain-rate tensor \mbox{$v^{\rm c}_{\alpha\beta} = (\partial_\alpha v^{\rm c}_\beta + \partial_\beta v^{\rm c}_\alpha)/2$} and where \mbox{$V_\alpha = v^{\rm c}_\alpha-v^{\rm f}_\alpha$}. The bulk viscosity $\bar\eta$ is a familiar term that is also present for a passive fluid, except that its microscopic origin is different for a tissue. The homeostatic pressure $P^{\rm c}_{\rm h}$ is a property specific for tissues that results from a balance of cell growth and cell death~\cite{basan2009homeostatic}. In addition to these terms, we have included additional terms in the isotropic stress: cell anisotropies can couple to the anisotropic stress, resulting in the term proportional to $\nu_0$. The terms proportional to $\nu_1$ and $\nu_2$ represent bioelectric and biohydraulic stresses induced by a coupling to the electric field or by a (relative) fluid flow, respectively. A similar expansion for the traceless anisotropic part of the stress tensor reads:
\begin{align}
\tilde\sigma^{\rm c}_{\alpha\beta}=2\eta \tilde v^{\rm c}_{\alpha\beta}+\zeta q_{\alpha\beta}-\nu_3 [E_\alpha p_\beta]_{\rm st} -\nu_4 [V_\alpha p_\beta]_{\rm st} \, . \label{eq_sigmaTilde}
\end{align}
where $\tilde v^{\rm c}_{\alpha\beta}$ is the traceless part of the cell strain-rate tensor, and we have moreover defined the symmetric traceless part of a dyadic product of vectors: $[A_\alpha B_\beta]_{\rm st}\equiv A_\alpha B_\beta + A_\beta B_\alpha - (2/3) A_\gamma B_\gamma \delta_{\alpha\beta}$. The coefficient $\eta$ is the usual shear viscosity while the other terms represent active couplings, and the terms proportional to $\nu_3$ and $\nu_4$ are the anisotropic counterpart of the terms proportional to $\nu_1$ and $\nu_2$ in Eq.~\eqref{eq_sigma}.
The active stress $\zeta q_{\alpha\beta}$ is a hallmark of active systems and shows the ability of cells to generate anisotropic stresses due to cell division or contraction of their cytoskeleton~\cite{simha2002hydrodynamic,bittig2008dynamics}. This active stress can be regulated by the cells and we therefore consider that it depends on the local pressure at linear order as:
\begin{align}
    \zeta = \zeta_0 - \zeta_1 (\sigma^{\rm c}+P^{\rm c}_{\rm h}) \, .
    \label{eq_activeStress}
\end{align}\label{eq_cell_stress}%
\end{subequations}%
We emphasize that the cell stress constitutive equations (Eqs.~\eqref{eq_sigma} to~\eqref{eq_activeStress}) reflect the effective viscous properties of tissues at long time as a consequence of cellular growth and death. We review the derivation of these constitutive equations in App.~\ref{sec_constitutive_equations}. A key feature of our work is that flows and electric fields do influence cell division and death, and therefore growth and shrinkage of the spheroid~\cite{blackiston2009bioelectric,sarkar2019field,duclut2019fluid}. 

The constitutive equation for the internal force density between cells and interstitial fluid including all the linear terms allowed by symmetry can be written as\footnote{\label{foot1}Note that additional terms must be added to Eqs.~\eqref{eq_momentum} and~\eqref{eq_eletricCurrent} for a system lacking spherical symmetry or for a system where the polarity is not purely radial with unit norm, see App.~\ref{sec_constitutive_not_spherical} for details.}
\begin{align}
f_\alpha\!=\!-\kappa(v^{\rm c}_\alpha \!-\! v^{\rm f}_\alpha) + \lambda_1 p_\alpha \!+ \lambda_2 E_\alpha 
\!+ \lambda_3 \partial_\beta q_{\alpha\beta} \, .
\label{eq_momentum}
\end{align}
The first term $-\kappa(v^{\rm c}_\alpha \!-\! v^{\rm f}_\alpha)$ accounts for the friction between the interstitial fluid and the cells and leads to Darcy's law~\cite{darcy1856fontaines} in the description of porous materials. The permeation coefficient can be estimated as $\kappa \simeq \eta^{\rm f}/a^2$, where $\eta^{\rm f}$ is the interstitial fluid viscosity and $a$ a typical interstitial distance.
The term $\lambda_1 p_\alpha$ accounts for active fluid pumping by the cells, while the terms proportional to $\lambda_2$ 
correspond to 
the electroosmotic contribution. The last term, proportional to $\lambda_3$, characterizes a differential pumping term due to the bending of the cells~\cite{ramaswamy2000nonequilibrium}.

To complete the tissue properties description, one finally needs to specify the constitutive equation for the electric current 
density%
\footnote{Note that additional terms must be added to Eqs.~\eqref{eq_momentum} and~\eqref{eq_eletricCurrent} for a system lacking spherical symmetry or for a system where the polarity is not purely radial with unit norm, see App.~\ref{sec_constitutive_not_spherical} for details.}
\begin{align}
j_\alpha\!=\!-\bar\kappa(v^{\rm c}_\alpha \!-\! v^{\rm f}_\alpha) + \Lambda_1 p_\alpha \!+ \Lambda_2 E_\alpha 
\!+ \Lambda_3 \partial_\beta q_{\alpha\beta} \, ,
\label{eq_eletricCurrent}
\end{align}
where $\Lambda_2$ 
is the electric conductivity of the tissue. The term proportional to $\bar\kappa$ characterizes the current due to the (relative) flow of ions between cells as a consequence of a reverse electroosmotic effect~\cite{kirby2013micro}. The coefficient~$\Lambda_1$ characterizes the contribution of ion pumping to the electric current, while the coefficient $\Lambda_3$ is an active flexoelectric coefficient. It indicates that a spatially nonuniform cell polarity orientation is obtained in response to an electric field, and has been shown to play a crucial role in the nucleation of a lumen in spherical cell aggregates~\cite{duclut2019fluid}.

    \subsection{Drain description}

The drain can be used to impose an external fluid flow or an external electric current in two equivalent ways: either by imposing directly a volumetric flow rate $Q^{\rm ext}$ and electric current $I^{\rm ext}$ through the drain, or, alternatively, a pressure difference $\Delta P = P^{\rm d}-P^{\rm ext}_2$ (with $P^{\rm ext}_2$ the pressure at the outer spheroid boundary and $P^{\rm d}$ the pressure at the outer end of the drain, see Fig.~\ref{fig_notation})
and an electric potential difference $\Delta U = U^{\rm d}-U^{\rm ext}_2$ (with $U^{\rm ext}_2$ the electric potential at outer spheroid boundary and $U^{\rm d}$ the electric potential at the outer end of the drain) can be applied.%

In both cases, the imposed external fluid flow and external electric current density at the lumen boundary read:
\begin{align}
    Q^{\rm ext} = 4\pi R_1^2 v^{\rm ext}_1 \, , \quad \text{and} \quad I^{\rm ext} = 4\pi R_1^2 j^{\rm ext}_1 \, ,
\end{align}
where $v^{\rm ext}_1$ and $j^{\rm ext}_1$ are the fluid velocity and electric current density at the boundary between the spheroid and the lumen.

In the following, we focus on the case where $Q^{\rm ext}$ and $I^{\rm ext}$ are imposed externally. The case of a pressure difference or electric potential difference imposed by the drain are discussed in App.~\ref{sec_other_ensembles}.

    \subsection{Continuity equations and boundary conditions}

If cell density and interstitial fluid density are equal and constant, which we assume in the following, then the total volume flux $v_\alpha = \phi v^{\rm c}_\alpha + (1-\phi)v^{\rm f}_\alpha$ is divergence-free: $\partial_\alpha v_\alpha = 0$
(see App.~\ref{sec_constitutive_equations} for details).
In the presence of the drain, which imposes a nonvanishing fluid velocity at the inner boundary of the spheroid, the integration of the total flow incompressibility in spherical coordinates yields a relation between the fluid velocity $v_r^{\rm f}$ and cell velocity $v_r^{\rm c}$ in the tissue, which reads:
\begin{align}
v_r^{\rm f} &= \frac{v^{\rm ext}_1}{1-\phi} \left(\frac{R_1}{r}\right)^2 -\frac{\phi}{1-\phi}v^c_r \, . \label{eq_vf_ext}
\end{align}

Similarly, charge conservation in the quasistatic limit $\partial_\alpha j_\alpha=0$ can be integrated in the case where an external current density $j^{\rm ext}_1$ is imposed on the inner boundary, yielding the current density $j_r = j^{\rm ext}_1 \left(R_1/r\right)^2$ throughout the tissue.

The spheroid is surrounded by an external fluid both inside (in the lumen), and outside. This fluid exerts a hydrostatic pressure on the tissue that is balanced by the tissue surface tension and by the total normal stress at the boundaries:
\begin{subequations}
\begin{align}
-\sigma^{\rm c}_{rr}(R_1) + P^{\rm f}(R_1) &= P^{\mathrm{ext}}_1-2\gamma_1/R_1  \, , \\
-\sigma^{\rm c}_{rr}(R_2) + P^{\rm f}(R_2) &= P^{\mathrm{ext}}_2+2\gamma_2/R_2 \, ,
\end{align}\label{eq_bc_stress}%
\end{subequations}%
where we have introduced the inner and outer tissue surface tensions $\gamma_1$ and  $\gamma_2$. Fluid exchange between the spheroid and the outside is driven by osmotic conditions: 
\begin{subequations}
\begin{align}
v^{\rm ext}_1 \!\!- \ddt{R_1} &= \!+ K_1 \left[(P^{\mathrm{ext}}_1- P^{\rm f}(R_1))-\Pi^{\mathrm{ext}}_1 \right]+J_{{\rm p},1} \, , \\
v^{\rm ext}_2 \!\!- \ddt{R_2} &= \!- K_2 \left[(P^{\mathrm{ext}}_2- P^{\rm f}(R_2))-\Pi^{\mathrm{ext}}_2 \right]-J_{{\rm p},2} \, .
\end{align} \label{eq_bc_osmo}%
\end{subequations}%
Here, $K_{1,2}$ are the permeability of the interfaces to water flow. The fluxes $J_{{\rm p},1}$ and $J_{{\rm p},2}$ can be nonzero as a result of active pumps and transporters that maintain an osmotic pressure difference, and act effectively as water pumps. We have also introduced $v^{\rm ext}_{2}$, the external flow imposed at outer spheroid boundary. Conservation of the volumetric flow directly yields $v^{\rm ext}_{2}=v^{\rm ext}_1 (R_1/R_2)^2$.

The normal velocity of the cells at the boundaries has to match the growth of the spheroid radii. 
An increased cell proliferation in a thin surface layer has been observed in growing spheroids~\cite{delarue2014stress,montel2011stress,delarue2013mechanical}. We thus allow for a thin surface layer of cells, both facing outside and to the lumen, to have a growth rate that differs from the bulk. The cell velocity boundary conditions then read:
\begin{subequations}
\begin{align}
    v_r^{\rm c}(R_1) = \ddt{R_1} + v_1 \, , \\
    v_r^{\rm c}(R_2) = \ddt{R_2} - v_2 \, ,
\end{align} \label{eq_bc_v}
\end{subequations}%
where $v_i=\delta k_i  n^{\rm c}_i h/n^{\rm c}$ with $h$ the thickness of the boundary layers, $n^{\rm c}_i$ and $\delta k_i$ the cell number density and the cell growth rate in the surface layers, respectively.

\section{Dynamics of the spheroid growth and orders of magnitude}

\begin{figure*}[t]
	\centering
    \subfigure[\label{fig_protocols_noExt_1}]
    {\includegraphics[width=0.32\linewidth]{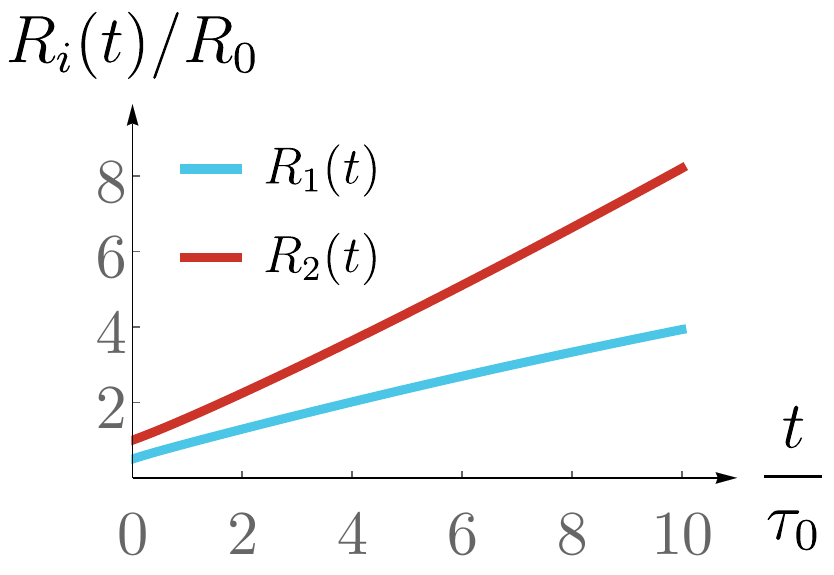}}
	\subfigure[\label{fig_protocols_sucess1}]
	{\includegraphics[width=0.32\linewidth]{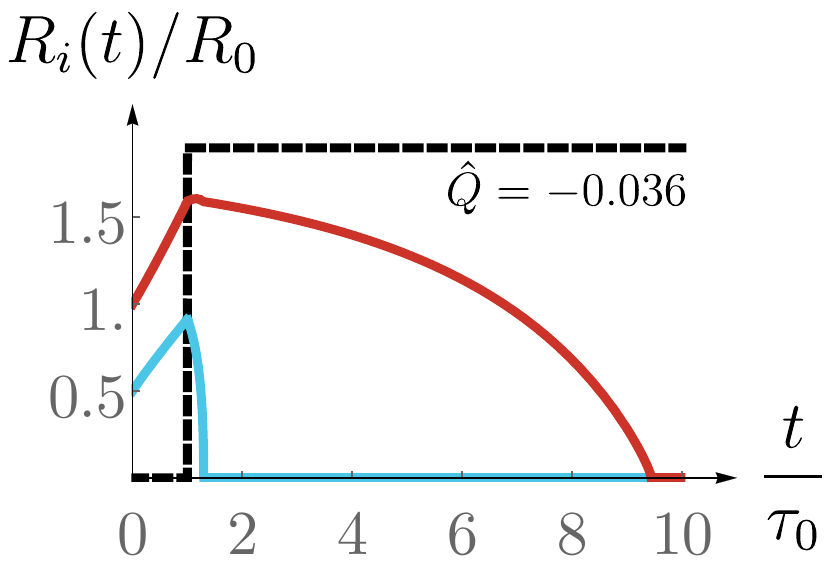}}
	\subfigure[\label{fig_protocols_sucess2}]
	{\includegraphics[width=0.32\linewidth]{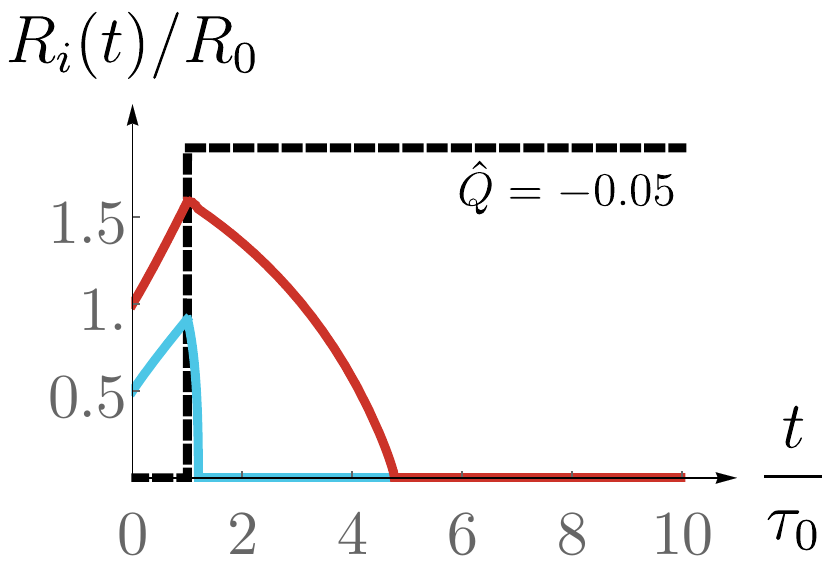}}
	\subfigure[\label{fig_protocols_sucess3}]
	{\includegraphics[width=0.32\linewidth]{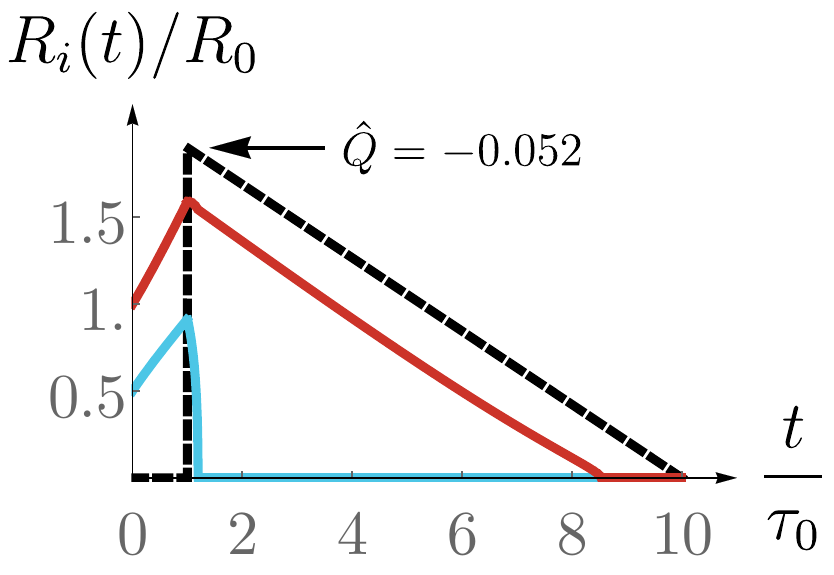}}
    \subfigure[\label{fig_protocols_unsucess1}]
    {\includegraphics[width=0.32\linewidth]{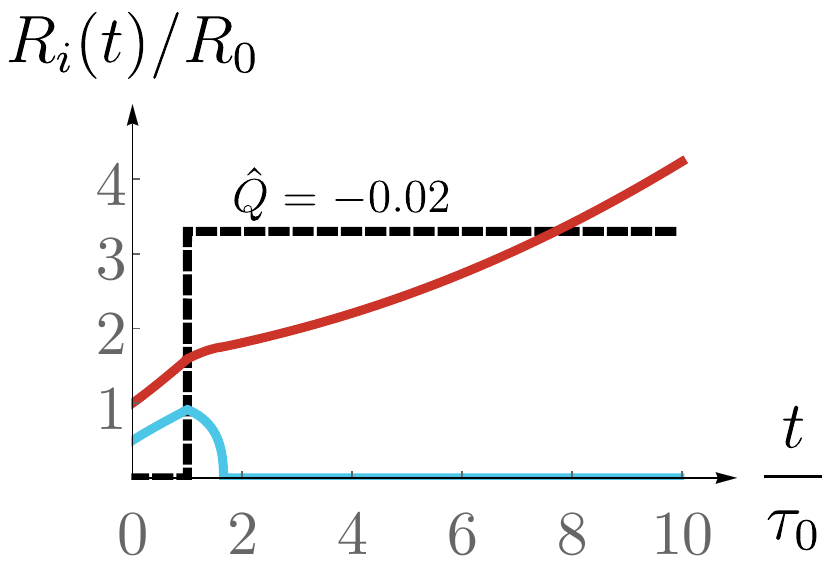}}
    \subfigure[\label{fig_protocols_unsucess2}]
    {\includegraphics[width=0.32\linewidth]{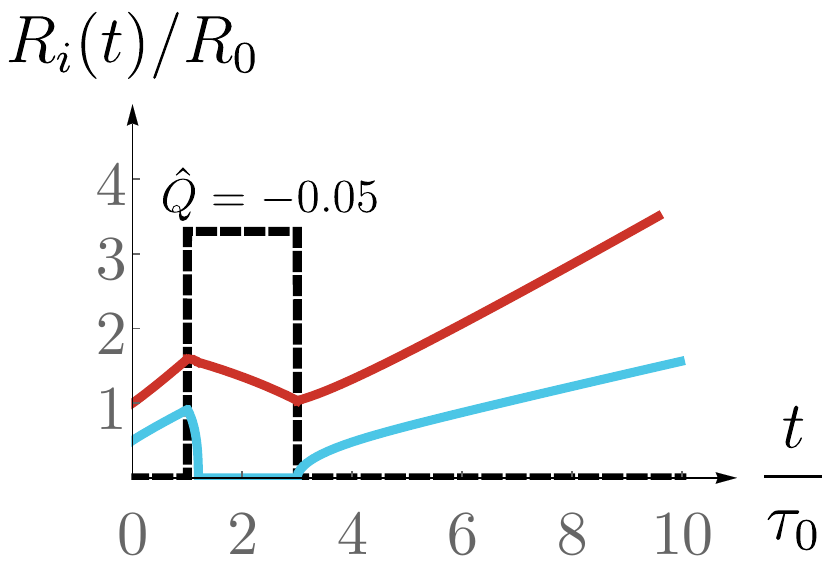}}

	\caption{Spheroid size $R_2$ (red curve) and lumen size $R_1$ (blue curve) normalized to a characteristic size $R_0$ as a function of the normalized time $\hat t = t/\tau_0$ for different external flow protocols \mbox{$\hat Q (\hat t) = Q^{\rm ext}(t)/Q_0$} (dashed black curve; its magnitude is not at scale). The definitions of $\tau_0,R_0$ and $Q_0$ are given in Eq.~\eqref{eq_normalization}. \textbf{(a)}~Spheroid growth without external intervention.  \textbf{(b)-(d)}~Successful protocols: the spheroid is suppressed. \textbf{(e)-(f)}~Unsuccessful protocols: the spheroid continues to grow. These plots are obtained by solving the dynamics equation~\eqref{eq_r1_Q} and~\eqref{eq_r2_Q} with the parameter values given in  Table~\ref{table_parameters_plots}.
	}
    \label{fig_protocols}
\end{figure*}

We have introduced in the previous section a model for a spherical spheroid with a drain. Solving force balance~\eqref{eq_forceBalance} together with the boundary conditions~\eqref{eq_bc_stress}-\eqref{eq_bc_v} then allow us to obtain the dynamics of the inner~$R_1(t)$ and outer~$R_2(t)$ radii of the spheroid in the quasistatic limit. We obtain two coupled nonlinear differential equations for the spheroid radii, Eqs.~\eqref{eq_r1_Q} and~\eqref{eq_r2_Q} (see App.~\ref{sec_derivation_radii}), which have been studied in Ref.~\cite{duclut2019fluid} in the absence of a drain. As we will see in the following, imposing an external fluid flow or electric current have dramatic consequences for the spheroid growth and can be used to control its size.

Before discussing how the presence of external fields modify the dynamics of a spheroid and which protocols can be used to control its growth, we first use our model to discuss the orders of magnitude of the external flux and electric current for which we expect a significant change in the spheroid dynamics. To obtain these estimates, we assume that the lumen size is small and use $R_1=0$ in Eq.~\eqref{eq_r2_Q} that describes the dynamics of the spheroid. We then compare in this equation the stresses generated by imposed flows or electric currents to the stresses stemming from internal activity in the absence of flows and currents. We find that the external volumetric flow required to significantly perturb the spheroid is of the order:
\begin{align}
    Q^{\rm ext} \simeq \sigma^{\rm typ} \frac{4\pi(1-\phi)}{\kappa^{\rm eff}} R_2 \, , \label{eq_orderMagnitude_Q}
\end{align}
where $\kappa^{\rm eff} = \kappa - \bar\kappa \lambda_2/\Lambda_2$
is an effective permeation coefficient and $\sigma^{\rm typ}$ is a typical scale of tissue stress in the absence of flows and currents. 
Note that the appearance of $\kappa^{\rm eff}$ in Eq.~\eqref{eq_orderMagnitude_Q} indicates that the finite bulk permeability of the spheroid governs the effects of the external flow imposed via the drain. Note however that for smaller spheroids of size $R_2\lesssim (1-\phi)/\kappa^{\rm eff} K_2\simeq 10-50~\mu$m, the flow-induced effects are not dominated by bulk permeation but rather by surface permeation and we have: $Q^{\rm ext} \simeq 4\pi K_2 \sigma^{\rm typ} R_2^2$.
We can estimate the external electric current required to perturb the spheroid:
\begin{align}
    I^{\rm ext} \simeq 4\pi \sigma^{\rm typ} \Lambda_2  R_2 / \lambda_2 \, . 
    \label{eq_orderMagnitude_I}
\end{align}

One can use experimental values and order-of-magnitude estimates of the parameters that appear in Eqs.~\eqref{eq_orderMagnitude_Q} and~\eqref{eq_orderMagnitude_I} (see Table~\ref{table_estimationParam}). We estimate that a volumetric flow \mbox{$Q^{\rm ext}\simeq 10^3-10^5~\mu$m$^3$/s} is sufficient to observe a significant change in the dynamics of a millimetre-sized spheroid, and this volumetric flow scales linearly with the size of the spheroid. Similarly, electric current of the order $I^{\rm ext} \simeq 1-100$~nA can perturb the dynamics of a millimetre-sized spheroid (see also Fig.~\ref{fig_orderMagnitude} in App.~\ref{sec_estimations}).

Externally imposed flows and electric currents could therefore be used to induce a change in the spheroid behavior. In the following sections, we focus in more details on how these external perturbations can be used to control the size of the spheroid.

\begin{table}[t!]
\setlength{\tabcolsep}{8pt}
\centering
    \begin{tabular}{ll||ll}
        \hline\hline
        \textbf{Parameters} & \textbf{Exp. values} & \textbf{Parameters} & \textbf{Estimations~\cite{sarkar2019field,duclut2019fluid}} \\
        \hline
        $\eta$ \cite{forgacs1998viscoelastic} & $10^{4}$~Pa$\cdot$s  & $\bar \kappa $ & $10^{3}$ A$\cdot$s$/$m$^{3}$  \\
        $\bar\eta$ \cite{montel2011stress} & $10^{9}$~Pa$\cdot$s & $\lambda_1$ & $-10^8$ N$/$m$^{3}$  \\
        $\gamma_{1,2}$ \cite{forgacs1998viscoelastic} & $10^{-3}$~N/m & $\lambda_2$ & $-10^6$ N$/$m$^{2}/$V \\
        $\kappa^{-1}$~\cite{netti2000role} & $10^{-13}$~m$^2$/Pa/s & $\lambda_3$ & $-10^3$ N$/$m$^{2}$  \\
        $\Pi_{1,2}^{\rm ext}$ 
        \cite{brace1977interaction}
        & $10^3$ Pa & $\Lambda_1$ &  $1$ A$/$m$^{2}$  \\
        $v_{1,2}$ \cite{delarue2014stress} & $10^{-10}$~m/s & $\Lambda_2$ & $10^{-2}$ A$/$V/m \\
        $P^{\rm c}_{\rm h}$  \cite{montel2011stress} & $-
        10^{3}$~Pa & $\Lambda_3$ & $10^{-5}$ A$/$m  \\
        $\zeta_0$ \cite{delarue2014stress} & $10^{3}$~Pa & $\nu_1$ & $10^1$ N/m/V  \\
        $\zeta_1$ \cite{delarue2014stress} & $-10^{-1}$ & $\nu_2$ & $10^8$ Pa$\cdot$s/m \\
        $R_{\rm d}$ & $5\, 10^{-6}$ m & $\nu_3$ & $1$ N/m/V  \\
        &  & $\nu_4$ & $10^7$ Pa$\cdot$s/m  \\
        &  & $K_{1,2}$ & $10^{-10}$~m/Pa/s   \\        
        & &$J_{{\rm p},1,2}$ & $10^{-11}$~m/s \\
        & & $1-\phi$ & $10^{-1}$  \\
        & & $\nu_0$ & $1$  \\
        \hline\hline
    \end{tabular}
    \caption{Experimental values and references (left columns) and  estimated values (right columns) of the phenomenological parameters of the model appearing in the constitutive equations.}
    \label{table_estimationParam}
\end{table}

\section{Hydraulic and electric control of the size of a spheroid}

	\subsection{Examples of protocols for spheroid suppression}

Using our model and solving numerically Eqs.~\eqref{eq_r1_Q} and~\eqref{eq_r2_Q} (see App.~\ref{sec_numerics}), we can analyse various protocols for the suppression of a spheroid. The dynamics of the spheroid and its lumen, $R_2(t)$ and $R_1(t)$, is therefore studied for different external flow protocols $Q^{\rm ext}(t)$ and electric current protocols $I^{\rm ext}(t)$. To keep the discussion as general as possible, we introduce dimensionless quantities: a dimensionless radius $r_2(\hat t)= R_2(t)/ R_0$, time $\hat t=t/\tau_0$, external volumetric flow $\hat Q = Q^{\rm ext}/Q_0$ and electric current $\hat I = I^{\rm ext}/I_0$, where we have defined:
\begin{align}
    R_0 = K_2 \bar \eta \, , &\quad \tau_0 = \bar\eta/|P^{\rm eff}_2| \, , \quad 
    Q_0 = 4\pi \bar\eta^2 K_2^3 |P^{\rm eff}_2| \, , \quad I_0 = 4\pi \Lambda_2 \bar\eta K_2 |P^{\rm eff}_2|/|\lambda_2| \, .
\label{eq_normalization}
\end{align}
Note that the effective pressure $P^{\rm eff}_2$ introduced above is a modification of the homeostatic pressure~$P^{\rm c}_{\rm h}$ by the external osmotic pressure and by electric and active contributions. Its expression can be found in Table~\ref{table_lookup} and in App.~\ref{sec_derivation_radii}.

We display in Fig.~\ref{fig_protocols} different protocols that can be applied to a growing spheroid in order to suppress it. If the external flow magnitude is sufficiently large and if it is applied long enough, the interventions are successful and lead to the spheroid suppression. Figure~\ref{fig_protocols_sucess1} shows an example of a successful suppression of the spheroid as a sufficiently strong flow has been applied until the spheroid is suppressed. A larger flow magnitude leads to a faster spheroid suppression (Fig.~\ref{fig_protocols_sucess2}). We also show in Fig.~\ref{fig_protocols_sucess3} an example of successful suppression protocol for which the magnitude of the flow is lowered as the spheroid size decreases.

\begin{figure*}[t]
	\centering
    \subfigure[\label{fig_protocols_noExt_2}]
    {\includegraphics[width=0.32\linewidth]{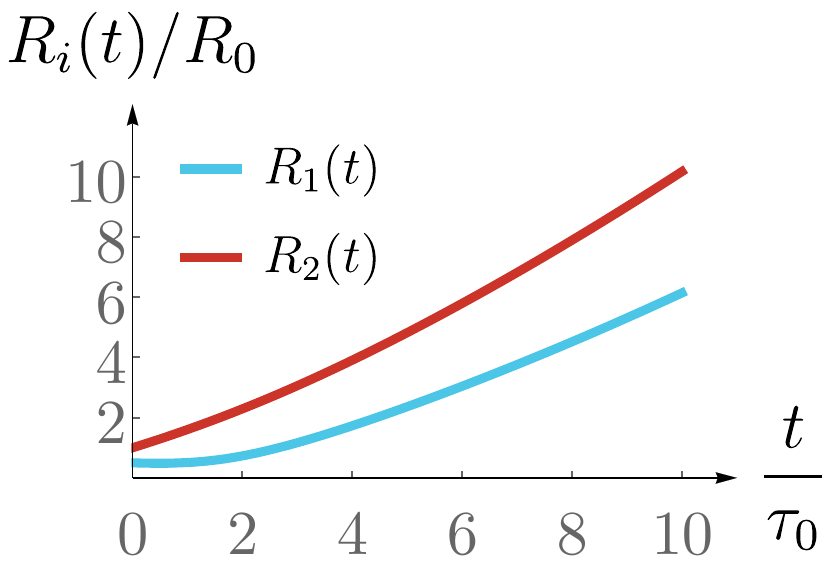}}
    \subfigure[\label{fig_protocols_I}]
    {\includegraphics[width=0.32\linewidth]{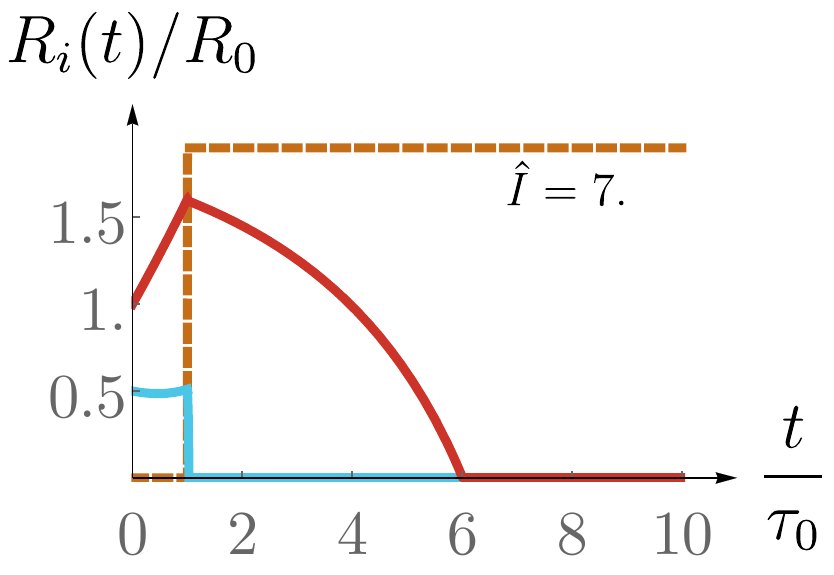}}
    \subfigure[\label{fig_protocols_IQ}]
    {\includegraphics[width=0.32\linewidth]{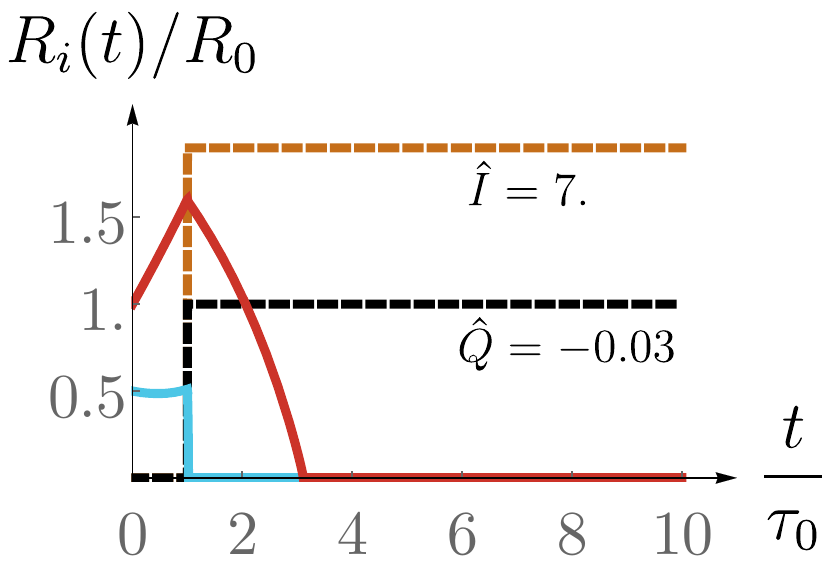}}
	\caption{Dimensionless spheroid size $R_2/R_0$ (red curve) and lumen size $R_1/R_0$ (blue curve) as a function of dimensionless time $\hat t = t/\tau_0$  in the presence of an imposed electric current \mbox{$\hat I (\hat t) = I^{\rm ext}(t)/I_0$} and imposed flow \mbox{$\hat Q (\hat t) = Q^{\rm ext}(t)/Q_0$} (dashed brown and black curve; their magnitude is not at scale). The definitions of $\tau_0,R_0, Q_0$ and $I_0$ are given in Eq.~\eqref{eq_normalization}. \textbf{(a)} Spheroid growth without external intervention. \textbf{(b)} Successful protocol with external electric field. \textbf{(c)} Successful protocol with external electric field and flow. Parameter values used for these plots can be found in Table~\ref{table_parameters_plots}.
	}
    \label{fig_protocols_current}
\end{figure*}

Conversely, if the magnitude of the imposed external flow is not sufficient, or if the duration of the flow is too short, the protocol can be unsuccessful in suppressing the spheroid. The spheroid may only have a slower growth rate (Fig.~\ref{fig_protocols_unsucess1}) or shrink significantly but resume growing as soon as the external flow is turned off (Fig.~\ref{fig_protocols_unsucess2}). Note that for any spheroid size, there exists a critical flow magnitude $|Q_c|$ such that imposing a steady flow at magnitude \mbox{$|Q^{\rm ext}|>|Q_c|$} will eventually lead to the spheroid suppression, while \mbox{$|Q^{\rm ext}|<|Q_c|$} will be unsuccessful. Figure~\ref{fig_protocols_sucess1} shows a protocol with \mbox{$|Q^{\rm ext}|\simeq|Q_c|$}.

In the examples above we have focused on the case where an external volumetric flow is imposed. The same analysis and similar procedures can be used in the case where an external current $I^{\rm ext}(t)$ is imposed (see Fig.~\ref{fig_protocols_current}). We show in Fig.~\ref{fig_protocols_I} a protocol that leads to the suppression of the spheroid. Note that an increased shrinking is obtained if one applies simultaneously an electric current and an external flow, as both effects are additive. Such intervention is displayed in Fig.~\ref{fig_protocols_IQ} where we observe a faster shrinking due to external flow in addition to the application of an electric current.

Importantly, we emphasize that the protocols we are discussing here are slow and take place on long time scales: using parameters values displayed in Table~\ref{table_parameters_plots}, the examples shown in Figs.~\ref{fig_protocols} and~\ref{fig_protocols_current} correspond to 
$R_0\simeq 1$~cm and $\tau_0\simeq 10$~days. This shows that the suppression of the spheroid requires a slow and steady flow or the application of a small electric current for several weeks.

\begin{figure*}[t]
	\centering
    \subfigure[\label{fig_phasePortrait_noExt}]
    {\includegraphics[width=0.32\linewidth]{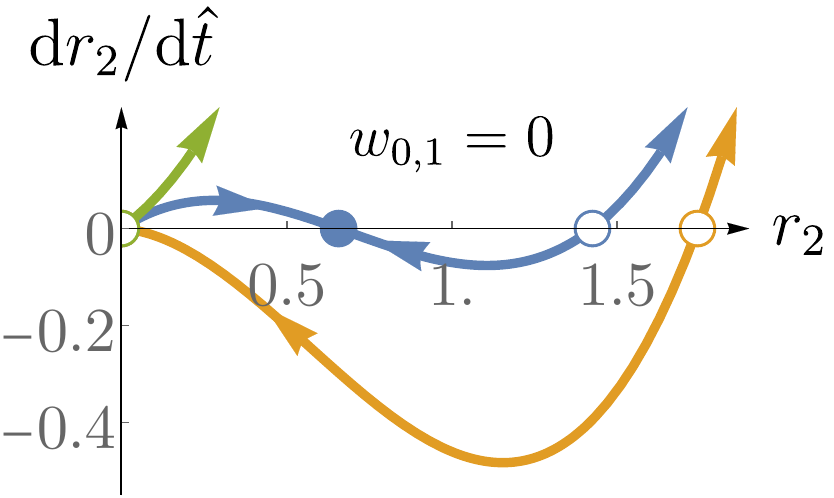}}
    \subfigure[\label{fig_phasePortrait_w0}]
    {\includegraphics[width=0.32\linewidth]{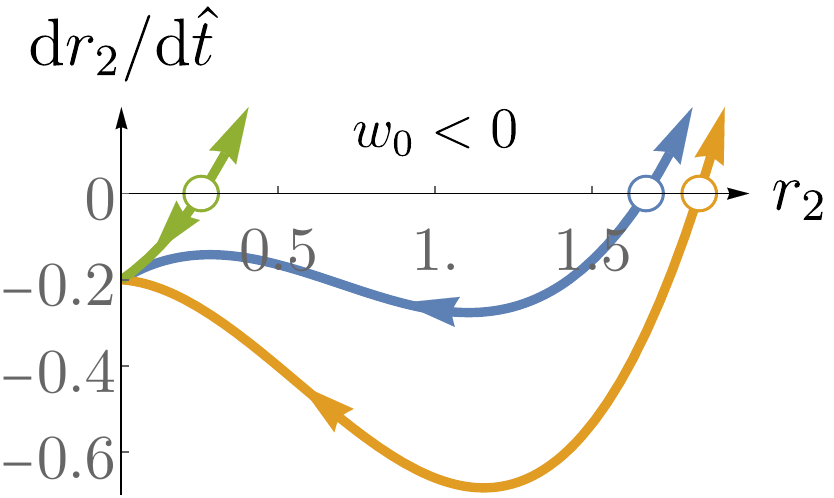}}
    \subfigure[\label{fig_phasePortrait_w1}]
    {\includegraphics[width=0.32\linewidth]{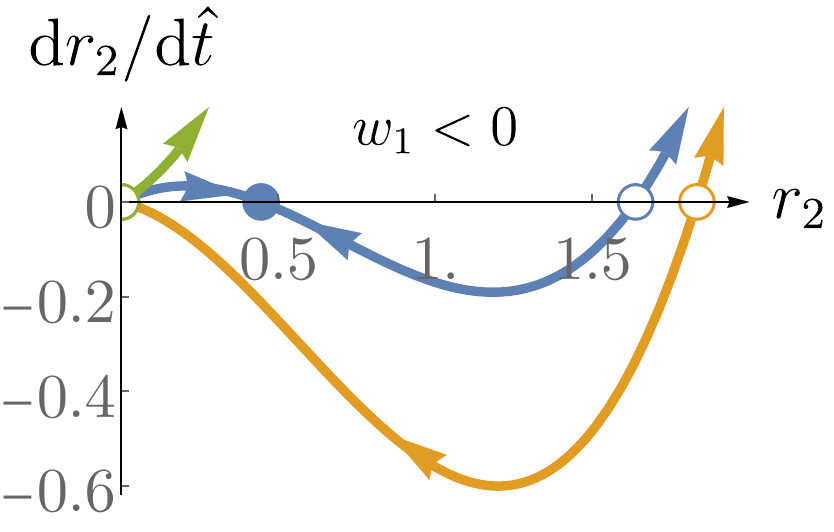}}
	\caption{Phase portraits $({\rm d}r_2/{\rm d}\hat t,r_2)$ of a spheroid as given by Eq.~\eqref{eq_r2_main}. \textbf{(a)} Example of phase portraits of spheroids that can grow in the absence of a drain. 
	\textbf{(b)} Example of the effect of a negative value of the parameter $w_0$ on the phase portraits. 
	\textbf{(c)} Example of a negative value of the parameter $w_1$ on the phase portraits. 
	Parameters (in the order green, blue, orange): $\hat v_2=\{0.5, 0.45, -0.05\}$, $\delta_2=\{-1,1,1\}$ and $\hat\lambda=\{2,1.92,2.36\}$. Drain parameters: plot (a), $w_0=w_1=0$; plot (b) $w_0=-0.2, w_1=0$; and plot (c) $w_0=0, w_1=-0.1$. Note that we have considered only the numerator of Eq.~\eqref{eq_r2_main} to draw these phase portraits as the denominator is always positive.
	}
    \label{fig_phasePortrait}
\end{figure*}

	\subsection{Spheroid dynamics with external flows and currents}

We now discuss quantitatively how a drain can be used to control the size of a spheroid. To keep the discussion as simple as possible, we consider here the limit of a spheroid enclosing a small lumen (or lumenless).
In the limit of a small lumen compared to spheroid size $R_1 \ll R_2$, the differential equation that describes the dynamics of the spheroid shrinks to (see App.~\ref{sec_derivation_radii}):
\begin{align}
    \ddroit{r_2}{\hat t} &= \frac{w_0 + (\hat v_2 +w_1/2) r_2 -\delta_2 \, r_2^2+ \hat\lambda \, r_2^3/4}{r_2(3+ r_2)} \, ,\label{eq_r2_main}
\end{align}
where we use the dimensionless radius $r_2$ and time $\hat t$ introduced in the previous section. 
The dimensionless parameters $w_{0,1}$ describe the effects of an externally-imposed flow or electric current. If they are set to zero, Eq.~\eqref{eq_r2_main} reduces to the dynamics of a lumenless spheroid without drain as studied in Ref.~\cite{duclut2019fluid}. 
They are defined as $w_n = j_n \hat I + u_n\hat Q$ with $n=0,1$ and where $\hat Q$ and $\hat I$ are the dimensionless volumetric flow and electric current, respectively, as defined above. 
The effects of an imposed external flow are captured by the coefficients:
\begin{align}
	&u_1 \!=\! \frac{\kappa^{\rm eff}\bar\eta K_2^2}{1-\phi} \, , \! \quad u_0 \!=\!  1-\beta_{\rm u} \, . 
	\label{eq_def_u01}
\end{align}%
Here, $u_1$ describes a permeation effect due to the finite permeability of the spheroid to fluid flows and involves the ratio of the surface permeability $K_2$ and of the (effective) bulk permeability $\propto 1/\kappa^{\rm eff}$. The parameter $u_0$ includes direct effects of the external flow -- proportional to $Q^{\rm ext}/K_2$ in the dimensional equation corresponding to $u_0=1$ --
which reflects the spheroid volume change imposed by the external flow due to tissue incompressibility (see also App.~\ref{sec_derivation_radii}). We have furthermore defined
\begin{align}
&\beta_{{\rm u}} = \frac{K_2}{1-\phi} \left(3 \nu_2 - 4\nu_0\nu_4 + \frac{\bar\kappa}{\Lambda_2} (3 \nu_1 - 4\nu_0\nu_3) \right) \, ,
\label{eq_def_betau}
\end{align}
which represents the bioelectric and biohydraulic contribution. Indeed, this term is a sum of terms proportional to the parameters $\nu_i$, and thus stems from the coupling between the electric field (or the interstitial fluid flow) and the cell polarity that appears in the cell stress.

Imposing an external electric current $I^{\rm ext}$ also contributes to the spheroid growth control via the parameters:
\begin{align}
    &  j_1 = 1 \, , \quad j_0 = \beta_{\rm j} \, .
    \label{eq_def_j01}
\end{align}%
The coefficient $j_1=1$ corresponds to electroosmotic flow due to the imposed current. The coefficient $j_0$ accounts for bioelectric and biohydraulic contributions where
\begin{align}
\beta_{{\rm j}} = \frac{3\nu_1-4\nu_0\nu_3}{\lambda_2\bar\eta K_2} \, .
    \label{eq_def_betaj}
\end{align}
Note that since the coefficients $\beta_{\rm u,j}$ are a sum of phenomenological parameters for which we only have order-of-magnitude estimates, it is difficult to obtain a reliable estimate of their magnitude and even of their sign. However, estimating upper bounds for $\beta_{\rm u,j}$ suggests that these bioelectric and biohydraulic contributions are small compared to the other effects. A further discussion of these contributions requires precise estimates of the bioelectric and biohydraulic couplings that could be obtained from experiments on spheroids in presence of a drain.

In equation~\eqref{eq_r2_main}, we have used the definitions~\cite{duclut2019fluid}
\begin{align}
    & \hat \lambda  \!=\!   \frac{ \lambda^{\rm eff} \bar\eta K_2}{|P^{\rm eff}_2|} \, ,  \quad \delta_2 = \frac{P^{\rm eff}_{2}}{ |P^{\rm eff}_2|} \, ,  \quad  \hat v_2 = \frac{v^{\rm eff}_2}{K_2 |P^{\rm eff}_2|} \, , 
    \quad v^{\rm eff}_2 = 3 v_2-\frac{2(\gamma^{\rm app}_{2}+\gamma^{\rm app}_{0})}{\bar\eta} \,  .
     \label{eq_dimless_parameters_noDrain}
\end{align}
Here, $\lambda^{\rm eff}$ is an an effective pumping coefficient and $\gamma^{\rm app}_{0,2}$ are apparent tensions defined in App.~\ref{sec_derivation_radii}. The apparent surface tension $\gamma^{\rm app}_2$  is a modification of the tissue surface tension~$\gamma_2$ stemming from the flexoelectric term proportional to $\Lambda_3$. This flexoelectric contribution plays a crucial role in lumen nucleation~\cite{duclut2019fluid}. In the dynamics of the outer radius of the spheroid, its effect is however minor. Finally, the effective pumping $\lambda^{\rm eff}$ combines the active pumping $\lambda_1$ and an electric contribution $\Lambda_1 \lambda_2/\Lambda_2$ due to electroosmosis.
The parameters introduced above are summarized for convenience in Table~\ref{table_lookup}, and their corresponding values can be computed using Table~\ref{table_estimationParam}.

\begin{table}[t]
    \setlength{\tabcolsep}{8pt}
    \centering
    \resizebox{\textwidth}{!}{%
    \begin{tabular}{llll}
    \hline\hline
    \textbf{Parameter} & \textbf{Definition} & \textbf{Equation} & \textbf{Description}\\ 
    \hline
    %
    $\kappa^{\rm eff}$ & $\kappa - \bar\kappa \lambda_2/\Lambda_2$ & below Eq.~\eqref{eq_orderMagnitude_Q}  &  effective permeation coefficient \\
    $P^{\rm eff}_{1,2}$ & $\Pi_{1,2}^{\rm ext}- P_{\rm h}^{\rm c} - \frac{J_{{\rm p},1,2}}{K_{1,2}} - \frac{2}{3} \left(\zeta_0\nu_0+ \lambda_3 +\frac{\Lambda_3\lambda_2 + (3\nu_1/2 -  2\nu_0\nu_3)\Lambda_1}{\Lambda_2} \right)$ & Eq.~\eqref{eq_Peff}  & effective pressure \\
    $\gamma^{\rm app}_{1,2}$ & $\gamma_{1,2} \mp 4\nu_3\Lambda_3/\Lambda_2$ & Eq.~\eqref{eq_gammaEff}  & apparent surface tension \\
    $\gamma^{\rm app}_0$ & $\left( 3\nu_1/2 - 2(2+\nu_0)\nu_3 \right) (\Lambda_3/\Lambda_2)$ & Eq.~\eqref{eq_gammaEff}  & apparent surface tension \\   
    $\lambda^{\rm eff}$ & $\lambda_1- \Lambda_1 \lambda_2/\Lambda_2$ & Eq.~\eqref{eq_lambdaEff}  & effective pumping coefficient \\
    $R_0$ & $K_2 \bar \eta$ & Eq.~\eqref{eq_normalization}  & characteristic length \\
    $\tau_0$ & $\bar\eta/|P^{\rm eff}_2|$ & Eq.~\eqref{eq_normalization} & characteristic time \\
    $Q_0$ & $4\pi \bar\eta^2 K_2^3 |P^{\rm eff}_2|$ & Eq.~\eqref{eq_normalization} & characteristic volumetric flux \\
    $I_0$ & $4\pi \Lambda_2 \bar\eta K_2 |P^{\rm eff}_2|/|\lambda_2|$ & Eq.~\eqref{eq_normalization} & characteristic electric current \\
    $\delta_{1,2}$ & $P^{\rm eff}_{1,2}/|P^{\rm eff}_{2}|$ & Eq.~\eqref{eq_dimless_parameters_noDrain} & dimensionless effective pressure \\
    $\hat \lambda$ & $\lambda^{\rm eff} \bar\eta K_2/|P^{\rm eff}_{2}|$& Eq.~\eqref{eq_dimless_parameters_noDrain} & dimensionless effective pumping \\
    $\hat v_2$ & $\left(3 v_2-(2(\gamma^{\rm app}_{2}+\gamma^{\rm app}_{0}))/\bar\eta\right)/(K_2 |P^{\rm eff}_2|)$& Eq.~\eqref{eq_dimless_parameters_noDrain} & dimensionless effective velocity \\    
    $\hat Q$ & $Q^{\rm ext}/Q_0$ & below~Eq.~\eqref{eq_r2_main} & dimensionless volumetric flow  \\
    $\hat I$ & $I^{\rm ext}/I_0$ & below~Eq.~\eqref{eq_r2_main} &  dimensionless electric current  \\
    $w_{0,1}$ & $j_{0,1} \hat I + u_{0,1} \hat Q$ & below~Eq.~\eqref{eq_r2_main} & dimensionless external contributions \\
    $u_0$ & $1-\frac{K_2}{1-\phi} \left(3 \nu_2 - 4\nu_0\nu_4 + \frac{\bar\kappa}{\Lambda_2} (3 \nu_1 - 4\nu_0\nu_3) \right)$ & Eqs.~(\ref{eq_def_u01}-\ref{eq_def_betau}) & / \\
    $u_1$ & $\kappa^{\rm eff}\bar\eta K_2^2/(1-\phi)$ & Eq.~\eqref{eq_def_u01} & / \\
    $j_0$ & $(3\nu_1-4\nu_0\nu_3)/(\lambda_2\bar\eta K_2)$ & Eqs.~(\ref{eq_def_j01}-\ref{eq_def_betaj}) & / \\
    $j_1$ & $1$ & Eq.~\eqref{eq_def_j01} & / \\
    \hline\hline
    \end{tabular}%
    }
    \caption{Summary of the effective parameters introduced in the text, their definition and the corresponding equation in the text.}
    \label{table_lookup}
\end{table}

    \subsection{Control of the spheroid dynamics}

We use  Eq.~\eqref{eq_r2_main} to discuss how external currents and flows can be used to control the growth and shrinkage of a spheroid. Note that whether the spheroid grows (${\rm d}r_2/{\rm d}t>0$) or shrinks (${\rm d}r_2/{\rm d}t<0$) only depends on the numerator of Eq.~\eqref{eq_r2_main}.

Figure~\ref{fig_phasePortrait} shows phase-space trajectories of the outer radius $r_2$ for spheroids that are able to grow in the absence of a drain. Figure~\ref{fig_phasePortrait_noExt} displays the case without external flow or current. The following scenarios are possible: the spheroid may be growing with an unstable fixed point at $r_2=0$ (green curve); alternatively, there might be an additional stable fixed point corresponding to a steady-state, either with finite radius $r_2^*\neq0$ (blue curve), or with vanishing radius $r_2^*$ (orange curve). Imposing an external flow implies nonzero values of $w_0$ and $w_1$, which are negative in the case where fluid flows out of the lumen (as depicted in Fig.~\ref{fig_notation_out}). 
The case of a nonvanishing value of $w_0$ is displayed in Fig.~\ref{fig_phasePortrait_w0}: in this case, the trajectories are shifted downward by an amount $w_0$. In this case, even the the growing spheroid (green curve) has a critical radius below which it shrinks (empty green circle). The stable steady-state (blue curve) has moved to $r_2^*=0$.
Figure~\ref{fig_phasePortrait_w1} shows the effect of a nonvanishing value of $w_1$: the slope of the phase-space trajectories is modified. Note that this perturbation favors shrinking of the spheroid: the unstable critical radius below which the spheroid shrinks (empty circles along the $x$ axis) is shifted to larger values compared to the case without external perturbation shown in Fig.~\ref{fig_phasePortrait_noExt}.

We can estimate the critical volumetric flow $Q_c$ needed to change a growing spheroid of size $R_2$ to a shrinking one. From Eq.~\eqref{eq_r2_main}, we obtain
\begin{align}
	Q_c = -\frac{4\pi K_2 R_2 (\bar\eta v^{\rm eff}_2-P^{\rm eff}_2 R_2 + \lambda^{\rm eff}R_2^2/4)}{1 + R_2 \kappa^{\rm eff} K_2/2(1-\phi) - \beta_{\rm u}} \, , \label{eq_Qc}
\end{align}
where the minus sign indicates that $Q^{\rm ext}<0$ is required to induce shrinkage (corresponding to flow out of the drain, see Fig.~\ref{fig_notation_out}).
We obtain a similar expression for the critical electric current that is required to induce shrinkage:
\begin{align}
	I_c = \frac{4\pi R_2 (\bar\eta v^{\rm eff}_2-P^{\rm eff}_2 R_2 + \lambda^{\rm eff}R_2^2/4)}{-R_2 \lambda_2/(2\Lambda_2) +  \bar\eta K_2 \beta_{\rm j}} \, .
\end{align}
Note that $\lambda_2<0$ and therefore $I_c$ is positive (corresponding to an electric current into the drain, see Fig.~\ref{fig_notation_in}). One can therefore always find a value of the externally applied flow or current to turn a growing spheroid into a shrinking one.

    \subsection{State diagrams of hydraulic control of spheroid growth}
    
\begin{figure*}[t]
	\centering
    \subfigure[\label{fig_stateDiagram_1}]
    {\includegraphics[width=0.31\linewidth]{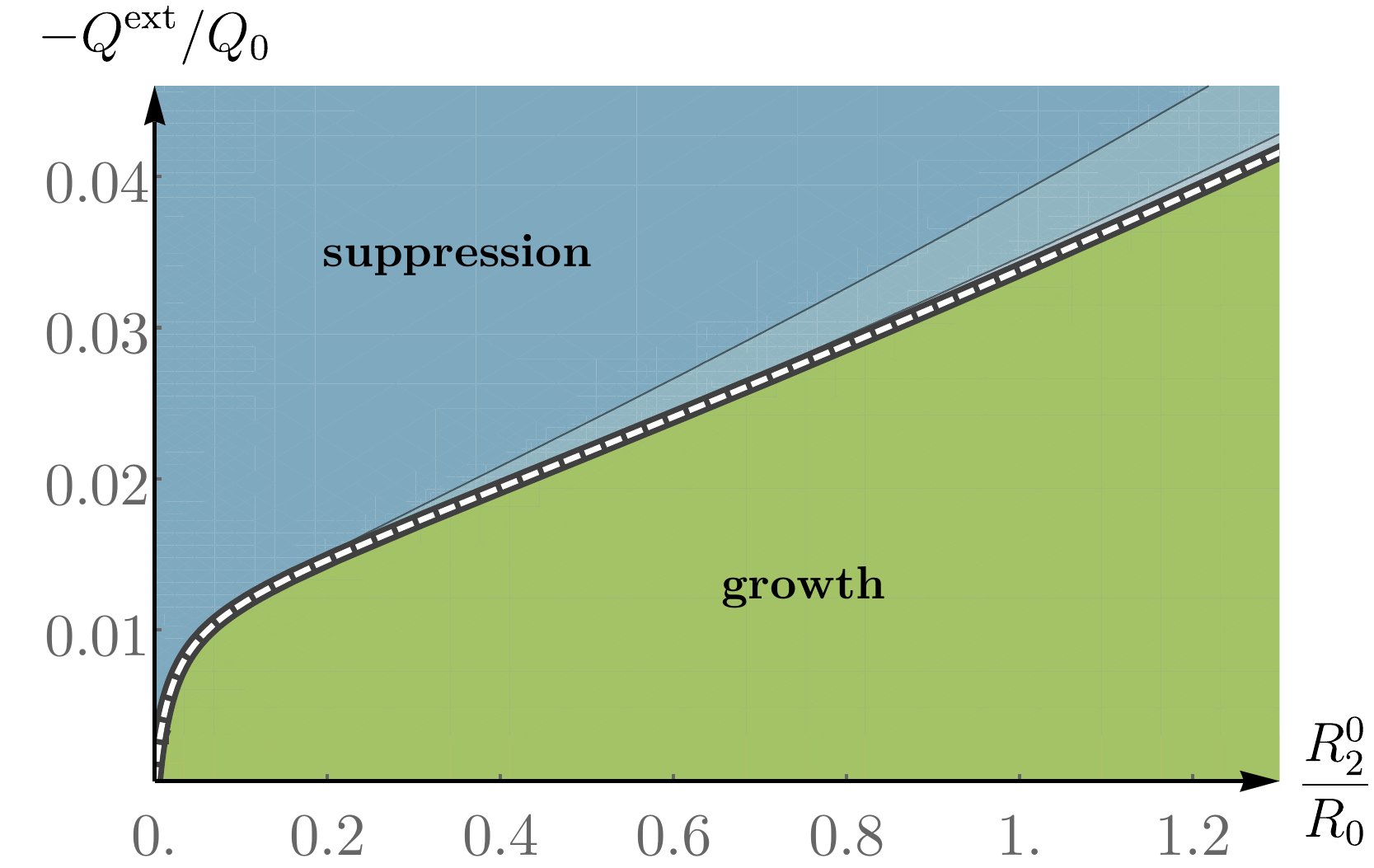}}
		\hfill
    \subfigure[\label{fig_stateDiagram_2}]
    {\includegraphics[width=0.31\linewidth]{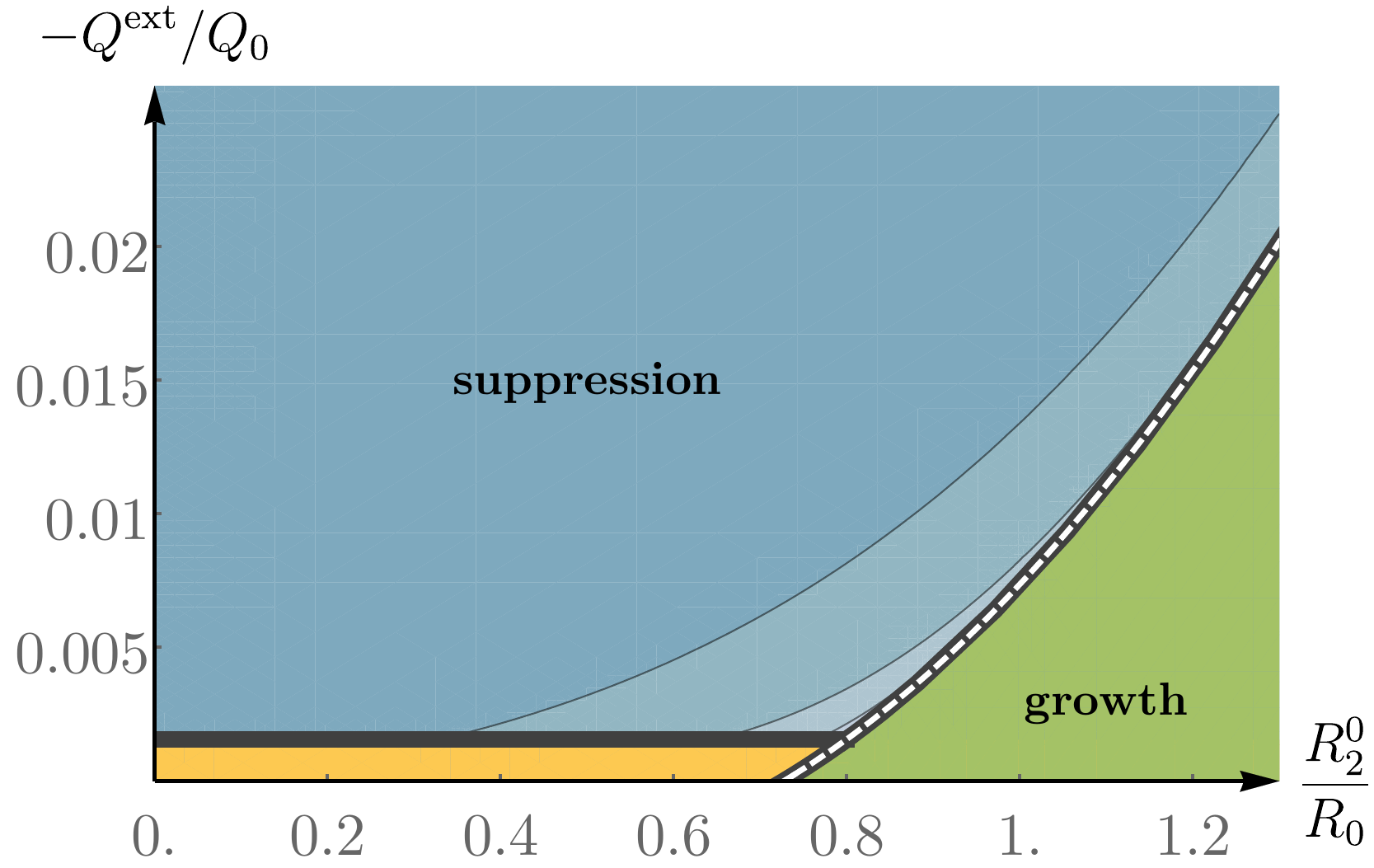}}
		\hfill
    \subfigure[\label{fig_stateDiagram_3}]
    {\includegraphics[width=0.345\linewidth]{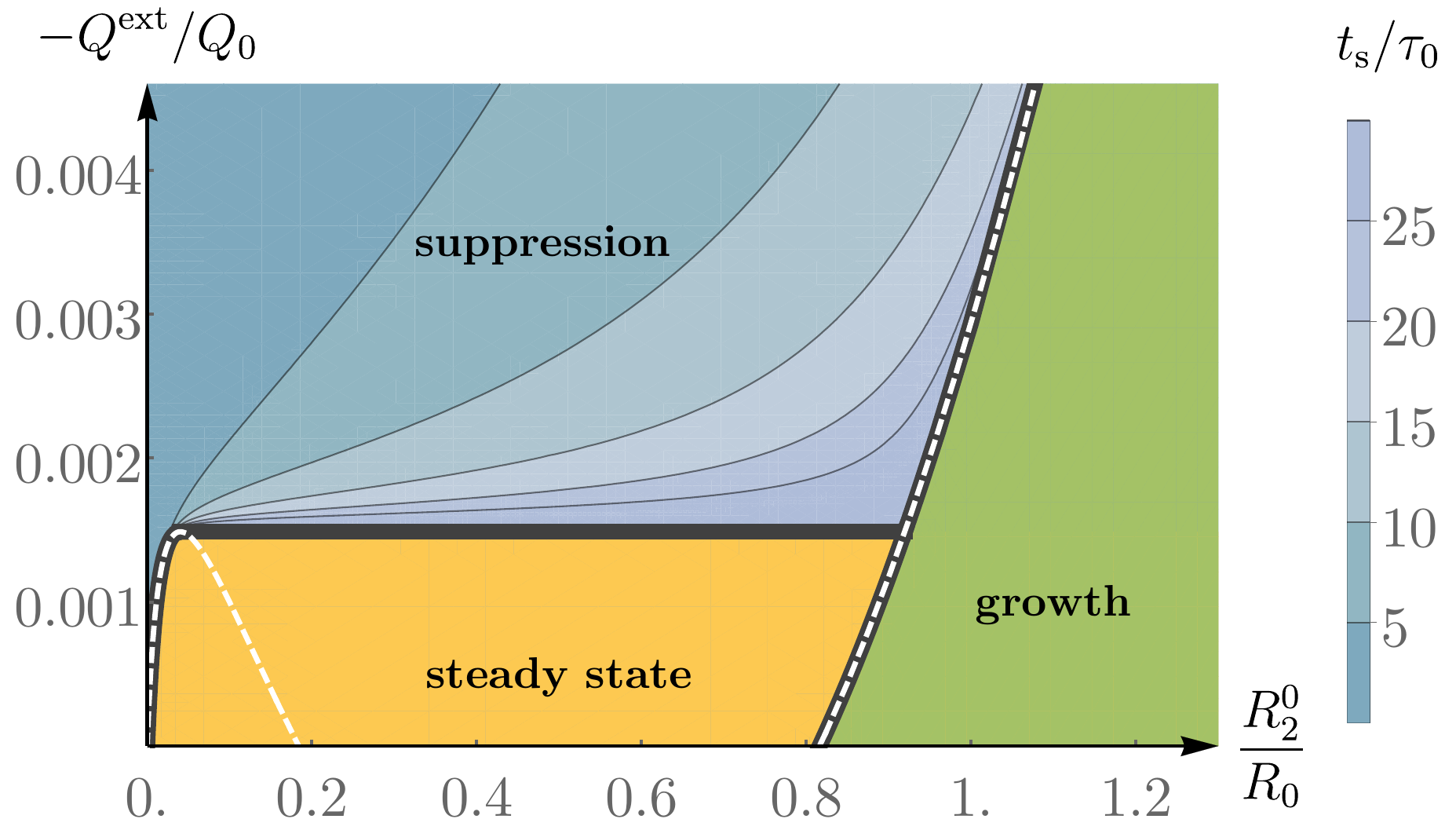}}
	\caption{%
	State diagrams of the long-time growth behavior of the spheroid as a function of its initial radius $R_2(t=0)=R_2^0$ and of the imposed volumetric flux $Q^{\rm ext}$. The state diagrams corresponds to a spheroid which, in the absence of the external flow and for vanishing lumen, is growing for all radii (a), is growing above a critical radius (b), reaches a steady-state radius or grows without bound (c). Shades of blue indicate the time for suppression $t_{\rm s}$ (see text). Note that all parameters are made dimensionless using the normalization given in Eq.~\eqref{eq_normalization}.
	These plots are obtained by solving the dynamics equations~\eqref{eq_r2_Q} with the parameter values given in  Table~\ref{table_parameters_plots} and with a fixed value of $R_1=2 R_{\rm d}$.
	}
    \label{fig_stateDiagrams}
\end{figure*}

We conclude this section by presenting state diagrams for a spheroid of initial size $R_2(t=0)=R_2^0$ to which a steady external flow $Q^{\rm ext}$ is imposed, see Fig.~\ref{fig_stateDiagrams}. We consider a spheroid with a drain inserted to the center but without lumen. This corresponds to choosing  $R_1$ equal to the outer radius of the drain. For simplicity, we fix $R_1=2R_{\rm d}$, where $R_{\rm d}$ is the inner drain radius. We again consider spheroids that can grow in the absence of a drain.
Figure~\ref{fig_stateDiagram_1} displays the state diagram for the case where the unperturbed spheroid is growing at all radii (corresponding to the green curve in Fig.~\ref{fig_phasePortrait_noExt}). Figure~\ref{fig_stateDiagram_2} shows the state diagram for a spheroid which is growing above a critical radius (corresponding to the yellow curve in Fig.~\ref{fig_phasePortrait_noExt}). Figure~\ref{fig_stateDiagram_3} is the state diagram for spheroids that, in the absence of a drain, either reach a steady-state radius or grow without bounds (corresponding to the blue curve in Fig.~\ref{fig_phasePortrait_noExt}). 

Three different regions exist in the diagrams shown in Fig.~\ref{fig_stateDiagrams}: (i)~a growth region (green), where the imposed flow is not sufficient to arrest growth. Note that for increasing values of $Q^{\rm ext}$, growth slows within this region; (ii)~a steady-state region (in orange), in which the spheroid reaches a finite size. Within this region a larger imposed flow drives the spheroid to smaller steady-state sizes; (iii)~a suppression region (in blue) where the spheroid shrinks until it is suppressed. We define that a spheroid has been suppressed when its size becomes smaller than a cut-off size $R_2=R_1+h_{\rm c}$, with $h_{\rm c}=10~\mu$m a typical cell thickness. 
In the region of spheroid suppression, the time $t_{\rm s}$ that it takes for the spheroid to be suppressed is indicated by shades of blue. For fixed initial spheroid size, the time for suppression decreases as the magnitude of the imposed flow increases. 

The solid black lines indicate boundaries between the different regions. 
Note that the line separating the steady-state region and the suppression region depends on the cut-off $h_{\rm c}$. The dashed white line indicates the nullcline ${\rm d}r_2/{\rm d}\hat t=0$ and gives the critical value of the external flow $Q_c$ required to switch from a growing to a shrinking spheroid (given by Eq.~\eqref{eq_Qc} for $R_1=0$). Note that below the nullcline the spheroids grow and above the nullcline they shrink.
Note that a similar analysis can be performed in the case of an external current $I^{\rm ext}$. In this case the same behaviors are found: growth, suppression and steady-state regions which depends on the initial spheroid size and on the magnitude of the imposed electric current.

\section{Conclusion}

Using a coarse-grained description of a tissue as an active material capable of exerting mechanical stresses, transporting ions and pumping fluid, we have proposed a technique to perturb tissues electrically and hydraulically. This technique relies on an imposed fluid flow or electric current source using a micrometric drain or electrode which can turn a growing spheroid to shrinkage. 

Our theory allows us to estimate orders of magnitude of the external flow and current that are required to suppress an initially growing spheroid. 
For small spheroids with radius of about a hundred micrometers, we show for instance that an electric current of a few nanoamperes or a volumetric flow of about $10^3~\mu$m$^3$/s could have a significant impact on the spheroid state (see  Fig.~\ref{fig_orderMagnitude} in App.~\ref{sec_estimations}). 
It means that even a passive drain connecting the inner part of the spheroid to the surrounding medium could already alter growth of small spheroids.
For larger aggregates, connecting the drain to a pump leads to spheroid suppression for any size, provided that the imposed external flow is sufficiently strong and maintained long enough 
(for instance, for spheroids of a few millimeters in radius an electric current of tens to hundreds of nanoamperes or a volumetric flow of about $10^4-10^5~\mu$m$^3$/s could be sufficient, see Fig.~\ref{fig_orderMagnitude}).
Spheroid suppression can be obtained also by imposing an electric current, and both fluid flow and electric current application could be combined to accelerate the process.

Our approach has also allowed us to characterize the different long-time states of a growing spheroid subject to an external perturbation. Depending on the magnitude of the hydraulic or electrical field, we have shown that the growth can be either slowed down or arrested (Fig.~\ref{fig_stateDiagrams}). In the latter case, the arrest of growth can lead to a steady state, or even to the shrinking and eventually the suppression of the spheroid. If the magnitude of the external field is above a threshold value, the suppression of the spheroid is always achieved, and stronger magnitudes then accelerate the process. Our coarse-grained description is generic, as it depends on effective tissue parameters but is insensitive to many details. Our approach does not distinguish between a suppressed state where all cells have disappeared and a state where a few cells still remain. This distinction depends on cellular details that we are not considering here.

We have also shown how different protocols can be used to obtain this suppression: for instance, a longer perturbation at a weaker magnitude is slower but as efficient as a shorter but stronger one; decreasing the magnitude of the flow or electric current as the size of the spheroid decreases can also be used to suppress a spheroid (see Fig.~\ref{fig_protocols} for examples). Interestingly, protocols that could lead to spheroid suppression need to be carried out over a sufficiently long time period. We estimate that the slow, steady flow or electric current that mediates the progressive suppression of a spheroid has to be maintained over days or weeks. This is in contrast with typical cancer ablation techniques~\cite{knavel2013tumor}, for which treatments are brief and intense. Radiofrequency ablation for instance requires the application of an alternating current with high frequency (up to 500 kHz) and high voltage (up to several kV) in order to heat the tissue~\cite{erez1980controlled}. Similarly, irreversible electroporation ablations use microsecond pulses of high electric potential (up to 3 kV)~\cite{miller2005cancer}. 

At the time scales considered in this manuscript, the case of a tumor is more complex than that of a spheroid. First, one has to compare the effect of the flux on the tumor with that on the surrounding healthy tissue. This question will be addressed in future work. 
Second, we have not explicitly considered the role of nutrient and oxygen transport. The larger division rate at the surface that we consider (see Eq.~\eqref{eq_bc_v}) could be related for example to oxygen gradient that has been observed experimentally~\cite{hirschhaeuser2010multicellular}. Furthermore, nutrient transport has been shown to influence growth in other contexts~\cite{wang2017shape}. 
Moreover, the geometry of spheroids could be more complex than a simple sphere, and a natural expansion of this current work will be the study of the stability of the spherical shape during the growth or shrinkage processes~\cite{martin2021viscocapillary}.
Last, during the suppression process, the escape of metastatic cells should be prevented. The signs of stresses that we find here corresponds to flows towards the sphere center. This suggests that escape of material might be hampered by imposed flows or currents, thereby providing a barrier against the escape of cells.

Our work is based on generic mechanical, hydraulic and electrical properties of tissues. It thus paves the way for novel experimental methods to control spheroid size. If applied to cancerous tissues, it could provide a means to influence cancerous tumor growth, arrest their proliferation and even suppress these tumors. Indeed, the hydraulic and electrical perturbations we have proposed here do not rely on identifying and hindering specific chemical pathways that are characteristic of cancerous tissue~\cite{hanahan2011hallmarks}, but rather on general, physical responses to an external field.

\subsection*{Acknowledgments}

We thank C.T. Lim and T.B. Saw for useful discussions. 

\clearpage

\appendix

\section{Tissues as active two-component fluids}
\label{sec_constitutive_equations}

    \subsection{Mass conservation in growing tissues}
    
In this work, we describe the tissue as a two-component system with (i) a cell phase that accounts for cells and the surrounding extra-cellular matrix, and (ii) the interstitial fluid that permeates the cell phase. Such a description has been introduced in Ref.~\cite{ranft2012tissue}. Here, we briefly review this approach to clarify the conceptual basis of Eq.~\eqref{eq_vf_ext} in the main text.

The cell phase is characterized by a cell mass $m^{\rm c}$, a cell number density $n^{\rm c}$, and a cell volume $\Omega^{\rm c}$, and we introduce similarly $m^{\rm f}$, $n^{\rm f}$ and $\Omega^{\rm f}$ for the interstitial fluid. The assumption that the cells and the fluid fill completely the space is written as:
\begin{align}
    n^{\rm c} \Omega^{\rm c} + n^{\rm f} \Omega^{\rm f} = 1 \, .
    \label{eq_volume_fill}
\end{align}
We use this assumption to define the cell volume fraction $\phi=n^{\rm c} \Omega^{\rm c}$ and the fluid volume fraction $n^{\rm f} \Omega^{\rm f}=1-\phi$.
Mass conservation in the tissue reads:
\begin{align}
    \partial_t \rho + \partial_\alpha j_\alpha = 0 \, ,
    \label{eq_mass_conservation}
\end{align}
where $\rho=m^{\rm c} n^{\rm c} + m^{\rm f} n^{\rm f}$ is the tissue density and $j_\alpha=m^{\rm c} n^{\rm c} v^{\rm c}_\alpha+ m^{\rm f} n^{\rm f} v^{\rm f}_\alpha$ is the total mass flux with $v^{\rm c,f}_\alpha$ the cell and interstitial fluid velocities. As a consequence of cell division and death, cell number is not conserved but obeys a continuity equation:
\begin{align}
\partial_t n^{\rm c} + \partial_\alpha(n^{\rm c} v^{\rm c}_\alpha)=n^{\rm c}(k_{\rm d}-k_{\rm a}) \, ,
\label{eq_continuity_cell}
\end{align}
where $k_{\rm d}$ and $k_{\rm a}$ are the rates of cell division and apoptosis respectively. Using mass conservation~\eqref{eq_mass_conservation}, the cell continuity equation~\eqref{eq_continuity_cell}, and assuming a constant fluid particle mass, we obtain the continuity equation for the fluid particle density~\cite{ranft2010fluidization}:
\begin{align}
\partial_t n^{\rm f} +\partial_\alpha(n^{\rm f} v^{\rm f}_\alpha)=-\frac{m^{\rm c}}{m^{\rm f}} n^{\rm c}(k_{\rm d}-k_{\rm a}) - \frac{n^{\rm c}}{m^{\rm f}} \ddroit{}{t} m^{\rm c}, \label{eq_continuity_fluid} 
\end{align}
where $(\ddt{})= \partial_t+v^{\rm c}_\gamma \partial_\gamma$ is the convected time derivative with respect to the cell flow. The above equation implies that a cell of mass $m^{\rm c}$ can be converted into $m^{\rm c}/m^{\rm f}$ fluid particles and vice versa when cells die or divide.

We define the total volume flux $v_\alpha = n^{\rm c} \Omega^{\rm c} v_\alpha^{\rm c} + n^{\rm f} \Omega^{\rm f} v_\alpha^{\rm f}$. Using Eqs.~\eqref{eq_volume_fill}, \eqref{eq_continuity_cell} and \eqref{eq_continuity_fluid} we obtain the following expression for its divergence~\cite{ranft2012tissue}:
\begin{align}
    \partial_\alpha v_\alpha =& \left( 1-\frac{\rho^{\rm c}}{\rho^{\rm f}} \right) \phi \left( k_{\rm d}-k_{\rm a} + \frac{1}{\Omega^{\rm c}}\ddroit{}{t}\Omega^{\rm c} \right) 
    - \frac{\phi}{\rho^{\rm f}} \ddroit{}{t} \rho^{\rm c} 
    -\frac{1-\phi}{\rho^{\rm f}} ( \partial_t + v_\alpha^{\rm f}\partial_\alpha) \rho^{\rm f} \, ,
\end{align}
where we have defined $\rho^{\rm c}=m^{\rm c}/\Omega^{\rm c}$ and $\rho^{\rm f}=m^{\rm f}/\Omega^{\rm f}$ the cell and fluid particle mass densities. Assuming that cell and fluid densities are constant and that $\rho^{\rm c}=\rho^{\rm f}$, the previous equation simplifies to yield $\partial_\alpha v_\alpha=0$~\cite{ranft2012tissue}. In the presence of the drain, which imposes a nonvanishing fluid velocity at the inner boundary of the spheroid, the integration of the total flow incompressibility in spherical coordinates yields Eq.~\eqref{eq_vf_ext} in the main text.

\subsection{Constitutive equations for the isotropic and anisotropic parts of the cell stress}
\label{sec_derivation_stress}

Following Refs.~\cite{ranft2010fluidization,ranft2012tissue,sarkar2019field}, we derive the constitutive equations for the cell stress $\sigma^{\rm c}_{\alpha\beta}$ for a permeated tissue in the presence of electric fields. 
We decompose the cell stress tensor into an isotropic contribution $\sigma^{\rm c}$ and a  traceless part $\tilde \sigma^{\rm c}_{\alpha\beta}$, such that $\sigma^{\rm c}_{\alpha\beta}=\tilde \sigma^{\rm c}_{\alpha\beta}+\sigma^{\rm c} \delta_{\alpha\beta}$.

We first discuss the isotropic cell stress.
Cell volume $\Omega^{\rm c}$ and cell volume fraction $\phi$ are in general functions of the isotropic cell stress $\sigma^{\rm c}$, the anisotropic cell stress $\tilde\sigma^{\rm c}_{\alpha\beta}$, the electric field $E_\alpha$, and the velocity difference $V_\alpha=v^{\rm c}_\alpha-v^{\rm f}_\alpha$. A general equation of state for the cell volume can therefore be written as:
\begin{align}
\Omega^{\rm c}=\Omega^{\rm c}(\sigma^{\rm c},q_{\alpha\beta}\tilde\sigma^{\rm c}_{\alpha\beta},p_\alpha E_\alpha, p_\alpha V_\alpha) \, , 
\label{eq_equation_of_state}
\end{align} 
and a similar expression for the cell volume fraction $\phi$.
Since $n^{\rm c}=\phi/\Omega^{\rm c}$, we can use the equation of state~\eqref{eq_equation_of_state} to write the time dependence of the cell number density:
\begin{align}
\frac{1}{n^{\rm c}} \ddroit{n^{\rm c}}{t} \!=\! -\frac{1}{\chi}\!\ddroit{\sigma^{\rm c}}{t}
- \frac{1}{\chi_0}\!\ddroit{(q_{\alpha\beta}\tilde\sigma^{\rm c}_{\alpha\beta})}{t}
-\frac{1}{\chi_1}\!\ddroit{(p_\alpha E_\alpha) }{t} -\frac{1}{\chi_2}\!\ddroit{(p_\alpha V_\alpha) }{t} \, , \label{eq_dt_equation_of_state} 
\end{align} 
where $\chi=n^{\rm c}(\partial n^{\rm c}/\partial \sigma^{\rm c})^{-1}$ and   $\chi_0=n^{\rm c}[\partial n^{\rm c}/\partial (q_{\alpha\beta}\tilde\sigma^{\rm c}_{\alpha\beta})]^{-1}$, denotes the isotropic and anisotropic compressibilities of the cells, and where we have defined the other compressibility coefficients as  $\chi_1=n^{\rm c}[\partial n^{\rm c}/\partial (p_\alpha E_\alpha)]^{-1}$, and $\chi_2=n^{\rm c}[\partial n^{\rm c}/\partial (p_\alpha V_\alpha)]^{-1}$. 
To write the constitutive equation for the isotropic stress in a closed form, we now eliminate $n^{\rm c}$. For this purpose the cell continuity equation~\eqref{eq_continuity_cell} can be rewritten as:
\begin{align}
\frac{1}{n^{\rm c}} \ddroit{n^{\rm c}}{t}= -v^{\rm c}_{\gamma\gamma}+k_{\rm d} - k_{\rm a} \, ,
\label{isostressrate}
\end{align}
and we specify a constitutive equation for the net growth rate of the tissue $k_{\rm d}-k_{\rm a}$. This growth rate in general depends  on the isotropic stress $\sigma^{\rm c}$ and may also depend on $q_{\alpha\beta}\tilde\sigma^{\rm c}_{\alpha\beta}$, $p_\alpha E_\alpha$ and $p_\alpha V_\alpha$. In the absence of anisotropic stresses, electric fields and flows, a constant cell density is achieved when cell death compensates cell division. The resulting isotropic cell pressure is the homeostatic pressure $P^{\rm c}_{\rm h}$. To linear order, the net growth rate near the homeostatic pressure reads~\cite{ranft2012tissue,sarkar2019field}: 
\begin{align}
k_{\rm d}-k_{\rm a}=\bar\eta^{-1}(P^{\rm c}_{\rm h} + \sigma^{\rm c} + \nu_0 \tilde\sigma^{\rm c}_{\alpha\beta}
q_{\alpha\beta} + \nu_1 p_\alpha E_\alpha+ \nu_2 p_\alpha V_\alpha) \, ,  \label{eq_cellGrowthRate}
\end{align}
where $\bar\eta$ is a constant and will be identified as the bulk viscosity in the following, $\nu_0$ is a dimensionless coefficient that takes into account the possible dependence of the growth rate on the anisotropic part of the stress, $\nu_1$ characterizes the influence of the electric field on the growth rate and $\nu_2$ is a coefficient accounting for the effects of the relative motion of the cells and the interstitial fluid to the growth rate.
Using Eqs.~\eqref{eq_dt_equation_of_state}-\eqref{eq_cellGrowthRate}, we obtain a general constitutive equation for the isotropic cell stress:
\begin{align}
    \begin{split}
    &\left(1+\tau\ddroit{}{t}\right)(\sigma^{\rm c} + P^{\rm c}_{\rm h}) + \nu_0 \left(1+\tau' \ddroit{}{t}\right)\tilde\sigma^{\rm c}_{\alpha\beta}q_{\alpha\beta} \\
    &+ \nu_1 \left(1+\tau_1\ddroit{}{t}\right)p_\alpha E_\alpha + \nu_2 \left(1+\tau_2\ddroit{}{t}\right) p_\alpha V_\alpha = \bar\eta v^{\rm c}_{\gamma\gamma} \, , 
    \end{split}\label{eq_isotropicStress}   
\end{align}
where $\tau=\bar\eta/\chi$ is and $\tau'=\bar\eta/\chi_0$ are the isotropic and anisotropic relaxation rates, $\tau_1=\bar\eta/\chi_1$ the relaxation rate associated with the electric field, $\tau_2=\bar\eta/\chi_2$ the relaxation rate arising from a velocity difference. At long times that we consider in the main text, we neglect the relaxation processes and Eq.~\eqref{eq_isotropicStress} reduces to
\begin{align}
\sigma^{\rm c} + P^{\rm c}_{\rm h} = \bar\eta 
v^{\rm c}_{\gamma\gamma} - \nu_0 \tilde\sigma^{\rm c}_{\alpha\beta}q_{\alpha\beta} - \nu_1 p_\alpha E_\alpha  - \nu_2 p_\alpha V_\alpha \, , 
\end{align}
which is Eq.~\eqref{eq_sigma} in the main text. In this long-time limit, cells are described as an active viscous fluid where $\bar\eta$ is revealed as an effective bulk viscosity due to cell division and death.

We now discuss the anisoptropic part of the cell stress $\tilde\sigma_{\alpha\beta}^{\rm c}$. We assume that the tissue behaves as an isotropic elastic material in the absence of cell division and apoptosis. When these events are considered, a reference state for the stress cannot be defined and we therefore express the changes of stress as a differential equation~\cite{ranft2012tissue}:
\begin{align}
    \corot \tilde\sigma^{\rm c}_{\alpha\beta} = 2\mu \tilde v^{\rm c}_{\alpha\beta}+ \corot \tilde\sigma^{\rm c, a}_{\alpha\beta} \, ,
\end{align}
where $({\rm D/D}t)\sigma^{\rm c}_{\alpha\beta} = \partial_t \sigma^{\rm c}_{\alpha\beta}+v^{\rm c}_\gamma \partial_\gamma \sigma^{\rm c}_{\alpha\beta} + \omega_{\alpha\gamma} \sigma^{\rm c}_{\gamma\beta} + \omega_{\beta\gamma} \sigma^{\rm c}_{\alpha\gamma}$ refers to the corotational time derivative with respect to the cell flow with $\omega_{\alpha\beta}=(\partial_\alpha v^{\rm c}_\beta-\partial_\beta v^{\rm c}_\alpha)/2$ the cell flow vorticity. This corotational derivative allows us to define a constitutive equation which does not depend on a frame of reference. 
In general, this term can depend on any of the traceless symmetric tensors in our model, such that at linear order we write:
\begin{align}
    \corot \tilde\sigma^{\rm c, a}_{\alpha\beta} =- \frac{1}{\tau_{\rm a}}\left(\tilde\sigma^{\rm c}_{\alpha\beta}-\zeta q_{\alpha\beta}+\nu_3 [E_\alpha p_\beta]_{\rm st} +\nu_4 [V_\alpha p_\beta]_{\rm st} \right) \, .
\end{align}
where $\tau_{\rm a}$ characterizes the anisotropic stress relaxation time. The constitutive equation for the anisotropic stress finally reads~\cite{ranft2010fluidization,delarue2014stress}:
\begin{align}
    \left(1+\tau_{\rm a}\corot \right) \tilde\sigma^{\rm c}_{\alpha\beta} = 2\eta \tilde v^{\rm c}_{\alpha\beta}+\zeta q_{\alpha\beta}-\nu_3 [E_\alpha p_\beta]_{\rm st} -\nu_4 [V_\alpha p_\beta]_{\rm st}  \, ,
\end{align}
where we have defined  the effective tissue shear viscosity $\eta=\mu\tau_{\rm a}$. In the main text we consider the long-time limit and the previous equation becomes:
\begin{align}
\tilde\sigma^{\rm c}_{\alpha\beta}=2\eta \tilde v^{\rm c}_{\alpha\beta}+\zeta q_{\alpha\beta}-\nu_3 [E_\alpha p_\beta]_{\rm st} -\nu_4 [V_\alpha p_\beta]_{\rm st} \, , 
\end{align}
which is Eq.~\eqref{eq_sigmaTilde} in the main text.

    \subsection{Constitutive equations for permeation and for the electric current density}
    
    \label{sec_constitutive_not_spherical}
    
Based on the symmetry of the system and assuming linear response, the constitutive equation for the internal force density reads:
\begin{align}
\begin{split}
f_\alpha\!&=\!-\tilde\kappa(v^{\rm c}_\alpha \!-\! v^{\rm f}_\alpha) -\tilde\kappa' q_{\alpha\beta}(v^{\rm c}_\beta \!-\! v^{\rm f}_\beta) + \tilde\lambda_1 p_\alpha + \tilde\lambda_1'q_{\alpha\beta} p_\beta \\
&\!+ \tilde\lambda_2 E_\alpha \!+ \tilde\lambda_2' q_{\alpha\beta}E_\beta \!+ \tilde\lambda_3 p_\alpha \partial_\beta p_\beta \!+ \tilde\lambda_3' p_\beta \partial_\beta p_\alpha \, .
\end{split}
\label{eq_momentum_appendix}
\end{align}
Note that the term $p_\alpha \partial_\beta p_\alpha=\partial_\beta (p_\alpha p_\alpha)/2$ vanishes as the polarity is a unit vector. Moreover, in the main text we consider a system with spherical symmetry and the polarity is assumed to be radial $\bm p= \bm{e}_r$, such that the nematic order parameter \mbox{$q_{\alpha\beta} = p_\alpha p_\beta - (1/3)\delta_{\alpha\beta}$} is diagonal in spherical coordinates with $q_{rr}=2/3$, $q_{\theta\theta}=-1/3$, $q_{\varphi\varphi}=-1/3$.
As a consequence, and since $p_\beta \partial_\beta p_\alpha=0$ for $\bm p= \bm{e}_r$, we obtain Eq.~\eqref{eq_momentum} in the main text with 
$\kappa=\tilde\kappa +2\tilde\kappa'/3$, $\lambda_1=\tilde\lambda_1 +2\tilde\lambda_1'/3$,  $\lambda_2=\tilde\lambda_2 +2\tilde\lambda_2'/3$, and $\lambda_3=\tilde\lambda_3$.

Similarly, the general form for the electric current density constitutive equation reads: 
\begin{align}
\begin{split}
j_\alpha\!&=\!-k(v^{\rm c}_\alpha \!-\! v^{\rm f}_\alpha) -k' q_{\alpha\beta}(v^{\rm c}_\beta \!-\! v^{\rm f}_\beta) + \tilde\Lambda_1 p_\alpha + \tilde\Lambda_1'q_{\alpha\beta} p_\beta \\
&\!+ \tilde\Lambda_2 E_\alpha \!+ \tilde\Lambda_2' q_{\alpha\beta}E_\beta \!+ \tilde\Lambda_3 p_\alpha \partial_\beta p_\beta \!+ \tilde\Lambda_3' p_\beta \partial_\beta p_\alpha \, .
\end{split}
\label{eq_eletricCurrent_appendix}
\end{align}
and it yields Eq.~\eqref{eq_eletricCurrent} in the main text with $\bar\kappa= k +2 k'/3$, $\Lambda_1=\tilde\Lambda_1 +2\tilde\Lambda_1'/3$,  $\Lambda_2=\tilde\Lambda_2 +2\tilde\Lambda_2'/3$, and $\Lambda_3=\tilde\Lambda_3$. for a system with spherical symmetry and with $\bm p = \bm e_r$.
For a system without spherical symmetry or with $\bm p \neq \bm e_r$, Eqs.~\eqref{eq_momentum_appendix} and~\eqref{eq_eletricCurrent_appendix} must be used.

\section{Dynamics of a spheroid in the presence of a drain}
\label{sec_derivation_radii}

The dynamical equations for the spheroid and lumen radii are obtained by first considering force balance~\eqref{eq_forceBalance} in spherical coordinates with boundary conditions~\eqref{eq_bc_v}. This yields the velocity profiles of cells and interstitial fluid. A differential equation for the radii dynamics $R_{1,2}(t)$ is then obtained from the normal stress balance~\eqref{eq_bc_stress} together with the permeation conditions at the boundary given by Eq.~\eqref{eq_bc_osmo}. Details on this computation in the case without a drain can be found in Ref.~\cite{duclut2019fluid}.
Including the external fields due to the presence of the drain, we obtain the following coupled differential equations for the dynamics of the radii $r_1$ and $r_2$ of the lumen and of the spheroid:
\begin{subequations}
\begin{align}
    \begin{split}
    &\chi \ddroit{r_1}{\hat t} + \frac{3(\hat V_1+\ddthat{r_1})r_1^2}{r_2^3-r_1^3} + \frac{3(\hat V_2-\ddthat{r_2}) r_2^2}{r_2^3-r_1^3} = \delta_1  \\
    &- \frac{2}{r_1}\left(\hat \gamma_1 -\hat \gamma_0 \frac{r_1 (r_1+r_2)}{r_1^2+r_2^2+r_1 r_2} \right)
    +(u_1\hat Q  +  \hat I) \frac{1}{r_1} \frac{(r_2-r_1)(r_1+2r_2)}{2(r_1^2+r_2^2+r_1 r_2)}  \\
    & + \frac{1}{r_1^2}\left(\chi \hat Q + u_2 \hat Q+ j_2\hat I  - (u_3\hat Q+ j_3 \hat I) \frac{(r_2-r_1)(2r_1+r_2)}{r_1^2+r_2^2+r_1 r_2}\right) \\
    & + \hat \lambda (r_2-r_1) \frac{r_1^2+2r_1 r_2+3 r_2^2}{4(r_1^2+r_2^2+r_1 r_2)} \, , \label{eq_r1_Q}
\end{split} \\
    \nonumber \\
\begin{split}
    &- \ddroit{r_2}{\hat t} +  \frac{3(\hat V_1+\ddthat{r_1})r_1^2}{r_2^3-r_1^3}+\frac{3(\hat V_2-\ddthat{r_2}) r_2^2}{r_2^3-r_1^3} = \delta_2 \\
    & + \frac{2}{r_2}\left(\hat \gamma_2+\hat\gamma_0 \frac{r_2 (r_1+r_2)}{r_1^2+r_2^2+r_1 r_2} \right)
    - (u_1\hat Q  +  \hat I) \frac{1}{r_2} \frac{(r_2-r_1)(2r_1+r_2)}{2(r_1^2+r_2^2+r_1 r_2)} \\
    & + \frac{1}{r_2^2} \left(-\hat Q + u_2 \hat Q+ j_2\hat I  + (u_3\hat Q+ j_3 \hat I) \frac{(r_2-r_1)(r_1+2r_2)}{r_1^2+r_2^2+r_1 r_2} \right)  \\
    & - \hat\lambda (r_2-r_1) \frac{3 r_1^2+2r_1 r_2 + r_2^2}{4(r_1^2+r_2^2+r_1 r_2)} \, , \label{eq_r2_Q}
\end{split}
\end{align}
\end{subequations}
where we have introduced dimensionless radii $r_i(\hat t)= R_i(t)/ R_0 $ with $R_0=K_2 \bar \eta$ and a dimensionless time $\hat t=t/\tau_0$ with $\tau_0=\bar\eta/|P^{\rm eff}_2|$. We have also introduced the dimensionless parameters:
\begin{align}
    &\hat \gamma_{0,1,2} \!=\! \frac{\gamma^{\rm app}_{0,1,2}}{\bar \eta K_2 |P^{\rm eff}_2|} \, , \quad \hat V_{1,2}=\frac{v_{1,2}}{K_2 |P^{\rm eff}_2|}  \, ,  \quad \delta_{1,2}  \!=\! \frac{P^{\rm eff}_{1,2}}{ |P^{\rm eff}_2|} \, , \quad
    \hat \lambda  \!=\!   \frac{ \lambda^{\rm eff} \bar\eta K_2 }{|P^{\rm eff}_2|}\, , \quad \chi\!=\! \frac{K_2}{K_1} \, ,
\label{eq_dimless_parameters}
\end{align}
where the effective parameters are defined as:
\begin{subequations}
\begin{align}
    & P^{\rm eff}_{1,2} = \Pi_{1,2}^{\rm ext}- P_{\rm h}^{\rm c} - \frac{J_{{\rm p},1,2}}{K_{1,2}} 
    - \frac{2}{3} \left(\zeta_0\nu_0+ \lambda_3 +\frac{\Lambda_3\lambda_2 + (3\nu_1/2 -  2\nu_0\nu_3)\Lambda_1}{\Lambda_2} \right)  , \label{eq_Peff}    \\
    & \gamma^{\rm app}_{1,2} = \gamma_{1,2} \mp 4\nu_3\Lambda_3/\Lambda_2 \, , \quad  \gamma^{\rm app}_0 = \left( 3\nu_1/2 - 2(2+\nu_0)\nu_3 \right) (\Lambda_3/\Lambda_2)  \, , \label{eq_gammaEff} \\
    &\lambda^{\rm eff}=\lambda_1- \Lambda_1 \lambda_2/\Lambda_2 \, . \label{eq_lambdaEff}
\end{align}
\end{subequations}
In addition to these terms, the presence of external flows and currents gives new dimensionless contributions which are defined as:
\begin{subequations}
\begin{align}
  & \hat Q = \frac{Q^{\rm ext}}{4\pi \bar\eta^2 K_2^3 |P^{\rm eff}_2|} \, , \quad
  \hat I = \frac{I^{\rm ext}|\lambda_2|}{4\pi \Lambda_2 \bar\eta K_2 |P^{\rm eff}_2|} \, , \quad
  u_1 = \frac{\kappa^{\rm eff}\bar\eta K_2^2}{1-\phi}  \, , \\
  & u_2 = \frac{K_2}{1-\phi}\left( \nu_2-4(\nu_0-1)\nu_4/3 -\frac{\bar\kappa}{\Lambda_2}(4(\nu_0-1)\nu_3/3-\nu_1) \right) \, , \\
  & u_3 = \frac{K_2}{1-\phi}\left( \nu_2 -2(1+2\nu_0)\nu_4/3 -\frac{\bar\kappa}{\Lambda_2}(2(1+2\nu_0)\nu_3/3-\nu_1) \right) \, , \\
  & j_2 = \frac{4(\nu_0-1)\nu_3/3-\nu_1}{\lambda_2 \bar\eta K_2 }  \, , \quad
  j_3 = \frac{2(1+2\nu_0)\nu_3/3-\nu_1}{\lambda_2 \bar\eta K_2} \, , \quad
\end{align}\label{eq_dimLess_all_uj}%
\end{subequations}%
where $\kappa^{\rm eff}= \kappa - \bar\kappa \lambda_2/\Lambda_2$ is an effective permeation coefficient.

To understand the effect of an external flux and an external electric current on the spheroid size, we can rewrite Eq.~\eqref{eq_r2_Q} in the form of an equation for the dynamics of the spheroid volume $V=4\pi R_2^3/3$. To simplify the analysis, we take a vanishing lumen size ($r_1=0$) and we consider only the terms that comes from the externally imposed current and flux. It yields:
\begin{align}
  \ddroit{V}{t} &= Q^{\rm ext} + Q^{\rm ext} \frac{K_2 \kappa^{\rm eff} R_2}{1-\phi} + I^{\rm ext} \frac{K_2 \lambda_2 R_2}{\Lambda_2}
  - Q^{\rm ext}(u_2+2u_3)
  - I^{\rm ext}\frac{\lambda_2 K_2}{\Lambda_2}(j_2+2j_3) \, ,
\end{align}
where we have reintroduced dimensional quantities to simplify the following discussion. The first term on the right-hand side of this equation is a direct consequence of imposing an external flow: since the tissue is incompressible, an imposed flow from the outside to the inside of the spheroid ($Q^{\rm ext}>0$) provokes an increase in volume, while an imposed flow from the inside of the spheroid to the outside ($Q^{\rm ext}<0$) shrinks its size. The second term is a signature of the finite permeability of the tissue, which acts as a porous medium, and the effect is therefore proportional to the ratio of the surface permeability $K_2$ and the bulk permeability $(\kappa^{\rm eff} R_2/(1-\phi))^{-1}$. The third term accounts for a similar phenomenon as the second one but due to the electroosmotically generated flow due to the imposed current.  The two last terms correspond to bioelectric and biohydraulic contributions: indeed, the terms $u_{2,3}$ and $j_{2,3}$ involve the parameters $\nu_i$ and thus stem from the coupling between the electric field (or the interstitial fluid flow) and the cell polarity that appears in the cell stress.

In the case of a small lumen compared to the spheroid size, one can consider the limit $R_1 \ll R_2$ to describe the spheroid dynamics. In this limit, we can in particular obtain from Eq.~\eqref{eq_r2_Q} an equation for the dynamics of $r_2$ only, which is given by Eq.~\eqref{eq_r2_main} in the main text. The parameters $\beta_{\rm u,j}$ introduced in the main text read $\beta_{\rm u} = u_2 + 2 u_3$ and $\beta_{\rm j} = j_2+ 2j_3$.

\section{Imposed pressure and electric potential difference} 
\label{sec_other_ensembles}

The main text describes the spheroid size control when an external volumetric flow $Q^{\rm ext}$ or an external current $I^{\rm ext}$ is imposed. Alternatively, a pressure difference  or an electric potential difference can be imposed between the outer part of the drain and the outer layer of the spheroid, which can be considered as the conjugate ensemble. In this appendix, we provide the additional equations required to discuss this situation.

    \subsection{Imposed pressure difference}

In order to compute the pressure difference $\Delta P=P^{\rm d} -P^{\rm ext}_2$, we decompose it as  $\Delta P = \Delta P_{\rm d1}+\Delta P_{12}$ where  $\Delta P_{12} = P^{\rm ext}_1 -P^{\rm ext}_2$ is the pressure difference across the spheroid and $\Delta P_{\rm d1} = P^{\rm d} -P^{\rm ext}_1$ is the pressure difference between the outer part of the drain and the lumen and reads:
\begin{align}
   \Delta P_{\rm d1} = Q^{\rm ext} (K_{\rm d} + K_{\rm lum}) \, , 
\end{align}
where $K_{\rm d, lum}$ are the hydraulic resistances of the drain and of the lumen, respectively. The drain hydraulic resistance is given by $K_{\rm d} = \frac{8\eta^{\rm f} L_{\rm d}}{\pi R_{\rm d}^4}$, where $\eta^{\rm f}\simeq 10$~mPa$\cdot$s is the viscosity of the interstitial fluid, $L_{\rm d},R_{\rm d}$ the width and radius of the drain.
    
The pressure difference $\Delta P_{12} = P^{\rm ext}_1 - P^{\rm ext}_2$ can then be computed using Eq.~\eqref{eq_bc_osmo}. In dimensionless form, it reads:
\begin{align}
    \Delta \hat P_{12} = \Delta \tilde \Pi + \hat Q \left( \frac{\chi}{r_1^2} + \frac{1}{r_2^2} \right) - (\chi \dot r_1 + \dot r_2) - \Delta \hat P^{\rm f} \, , \label{eq_deltaP}
\end{align}
where we use the short-hand notation $\dot r_{1,2} = \ddthat{r_{1,2}}$ and we have defined:
\begin{align}
\begin{split}
    &\Delta \hat P_{12} = \frac{P^{\rm ext}_1-P^{\rm ext}_2}{|P^{\rm eff}_2|} \, , \quad \Delta \hat P^{\rm f}= \frac{P^{\rm f}(r_1)-P^{\rm f}(r_2)}{|P^{\rm eff}_2|} \\
    &\Delta \tilde \Pi = \frac{\Pi_1-\Pi_2 - (J_1/K_1 - J_2/K_2)}{|P^{\rm eff}_2|} \, .
\end{split}
\end{align}
The dimensionless quantities $P_2^{\rm eff}$, $r_{1,2}$, $\chi$ and $\hat Q$ have been introduced in App.~\ref{sec_derivation_radii}. To compute the interstitial pressure difference $\Delta P^{\rm f}= P^{\rm f}(R_1)-P^{\rm f}(R_2)$, one can use the identity $\Delta P^{\rm f} =- \int_{R_1}^{R_2} \dd r \, \partial_r P^{\rm f}(r)$. Indeed, the gradient of the interstitial fluid pressure can be obtained using force balance~\eqref{eq_forceBalance_fluid} and the constitutive equations~\eqref{eq_momentum} and~\eqref{eq_eletricCurrent}. It reads:
\begin{align}
    \partial_{\hat r} \hat P^{\rm f}(\hat r) = u_1 \hat v(\hat r) - \hat \lambda - \frac{2 \hat \lambda_3}{\hat r} - \frac{u_1 \hat Q+\hat I}{\hat r^2}\, , \label{eq_gradPf}
\end{align}
where we have introduced, in addition to the dimensionless quantities already introduced in App.~\ref{sec_derivation_radii}, the dimensionless radius $\hat r = r/R_0$, cell velocity $\hat v = v^{\rm c}_r/(K_2|P^{\rm eff}_2|)$, and interstitial fluid pressure  $\hat P^{\rm f} = P^{\rm f}/|P^{\rm eff}_2|$ and the dimensionless parameter
$\hat \lambda_3 = (\lambda_3-\Lambda_3 \lambda_2/\Lambda_2)/|P^{\rm eff}_2|$.
The dimensionless cell velocity appearing in Eq.~\eqref{eq_gradPf} can be cast into the following form:
\begin{align}
\hat v(\hat r) = 
\frac{a}{\hat r^2} +
\frac{b}{\hat r} +
c
+ d \hat r
- \frac{\hat \lambda}{4} \hat r^2 \, , \label{eq_v_dimless}
\end{align}
where we have introduced the following $\hat r$-independent factors:
\begin{subequations}
\begin{align}
\begin{split}
a &= \frac{r_1^2 r_2^2}{r_{12}^2}\left(r_1 (\tilde v_2-\dot r_2) + r_2 (\tilde v_1+\dot r_1) \right)
- \frac{\hat \lambda r_1^3 r_2^3}{4 r_{12}^2} \\
& + \frac{\hat Q r_1 r_2}{r_{12}^2} \left( u_3(r_1+r_2) - \frac{u_1 r_1 r_2}{2} \right) + \frac{\hat I r_1 r_2}{r_{12}^2} \left( j_3(r_1+r_2) - \frac{ r_1 r_2}{2} \right) \, ,
\end{split} \\
b &= -u_3 \hat Q - j_3 \hat I \, , \quad
c = \frac{u_1 \hat Q}{2} + \frac{\hat I}{2} -  \frac{2 \hat \gamma_0}{3} \, ,\\
\begin{split}
d &= \frac{\hat\lambda}{4}\left(r_2 + \frac{r_1^3}{r_{12}^2} \right) - \frac{(\tilde v_1+\dot r_1)r_1^2 +(\tilde v_2-\dot r_2)r_2^2}{r_2^3-r_1^3}  \\
&\quad + \frac{\hat Q}{r_{12}^2} \left(u_3 - \frac{u_1(r_1+r_2)}{2} \right) + \frac{\hat I}{r_{12}^2} \left(j_3 - \frac{r_1+r_2}{2} \right) \, ,
\end{split}
%
\end{align}
\end{subequations}%
and the following dimensionless quantities (in addition to those already introduced in App.~\ref{sec_derivation_radii}):
\begin{align}
    r_{12}^2 &= r_1^2 + r_2^2 + r_1 r_2 \, , \quad
    \tilde v_{1,2} = \hat V_{1,2} \mp 2\hat \gamma_0/3 \, .
\end{align}

The dimensionless pressure difference~\eqref{eq_deltaP} can therefore be computed in terms of the dimensionless parameters of the model, although its expression is lengthy. One can however expand this expression in the limit of a thin spheroid (i.e $R_1\sim R_2$), yielding the expression:
\begin{align}
    \Delta \hat P_{12} = \frac{1}{|P^{\rm eff}_2|} \frac{2(\gamma_1+\gamma_2)}{R_2} + (r_2-r_1) A + \mathcal{O}\left( (r_2-r_1)^2 \right) \, ,
\end{align}
where the first term on the right-hand side is the Laplace contribution in the limit $R_1\to R_2$ and the dimensionless coefficient $A$ represents the first correction in term of the thickness $R_2-R_1$ and reads:
\begin{align}
\begin{split}
    A &=  \frac{u_1\left(\hat V_2-\chi\hat V_1 -\Delta \tilde \Pi \right)}{1+\chi} + \frac{2}{r_2} \left( \frac{u_1 (\hat \gamma_1 + \hat\gamma_2)}{1+\chi} +\hat \lambda_3   \right) + \frac{2\hat\gamma_1}{r_2^2}\\
    & \quad + \frac{\hat Q}{r_2^2}\left(\frac{2(u_2-u_3)}{r_2}-u_1 \right) + \frac{2\hat I (j_3-j_2)}{r_2^3} \, .
\end{split}
\end{align}

    \subsection{Imposed electric potential difference}

Similarly to the pressure, one can compute the electric potential difference $\Delta U=U^{\rm d} -U^{\rm ext}_2$ by decomposing it as  $\Delta U = \Delta U_{\rm d1}+\Delta U_{12}$ with $\Delta U_{12} = U^{\rm ext}_1 - U^{\rm ext}_2$ the electric potential difference across the spheroid and with $\Delta U_{\rm d1} = U^{\rm d} - U^{\rm ext}_1$ the electric potential difference across the drain. This difference reads
\begin{align}
    \Delta U_{\rm d1} = I^{\rm ext} (1/G_{\rm d}+ 1/G_{\rm lum}) \, ,
\end{align}
where $G_{\rm d, lum}$ are the electrical conductances of the drain and of the lumen. The electrical conductance of the drain is computed as $1/G_{\rm d} = \frac{\rho^{\rm f} L_{\rm d}}{\pi R_{\rm d}^2}$, where $\rho^{\rm f}\simeq 1~{\rm \Omega}\cdot$m is the resistivity of the interstitial fluid (taken to be that of salted water, whose conductivity is about \mbox{$G\simeq 1$ S/m}).

The electric potential difference across the spheroid $\Delta U_{12}$ can then be computed in the following way. From Eq.~\eqref{eq_eletricCurrent}, an expression for the electric field can be obtained, which can then be used to compute the electric potential difference using the identity \mbox{$\Delta U_{12}=\int_{R_1}^{R_2} \dd r \, E(r)$}. Defining the dimensionless electric field $\hat E = E_r \, \bar \eta K_2 \lambda_2/|P_2^{\rm eff}|$, we have:
\begin{align}
    \hat E (\hat r) = - \hat \Lambda_1 - \bar u_1 \hat v(\hat r) -\frac{2\hat \Lambda_3}{\hat r} - \frac{\bar u_1 \hat Q - \hat I}{\hat r^2} \, ,
\end{align}
where we have defined the dimensionless parameters:
\begin{align}
    \bar u_1 = \frac{\bar \eta K_2^2 \bar\kappa \lambda_2 }{(1-\phi)\Lambda_2} \, , \quad  \hat \Lambda_1 = \frac{\bar \eta K_2 \lambda_2 \Lambda_1}{\Lambda_2 |P_2^{\rm eff}|} \, , \quad \hat \Lambda_3 = \frac{\lambda_2 \Lambda_3}{\Lambda_2 |P_2^{\rm eff}|} \, ,
\end{align}
and where the expression for the dimensionless cell velocity $\hat v$ is displayed in Eq.~\eqref{eq_v_dimless}.

\begin{figure}[t]
	\centering
		\subfigure[]
    {\includegraphics[width=0.45\linewidth]{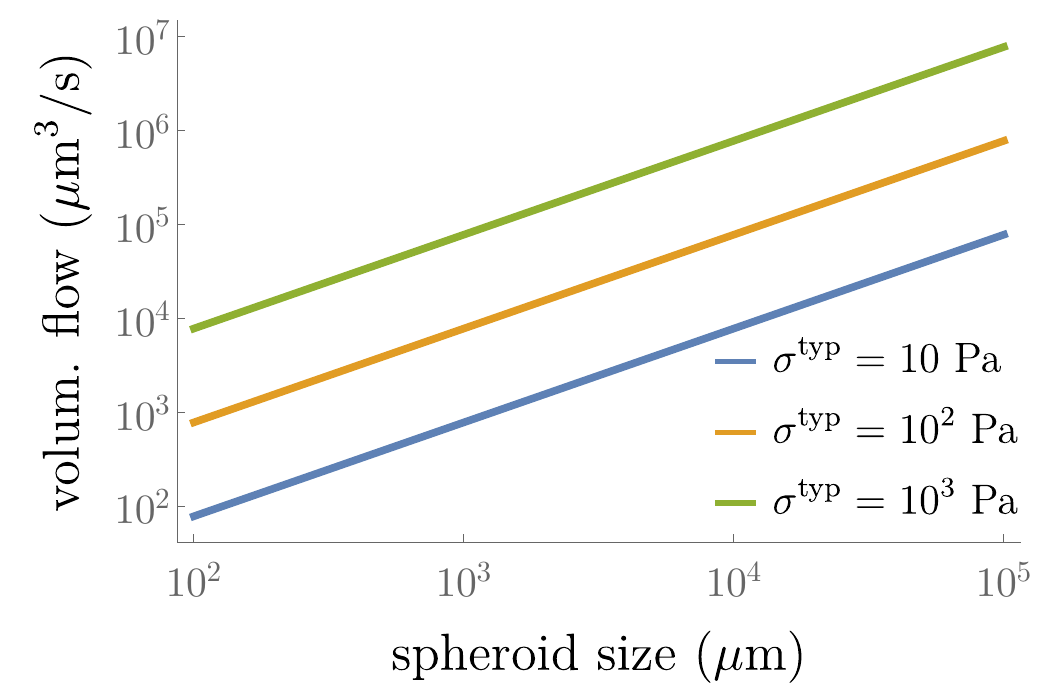}}
    \hfill
		\subfigure[]
		{\includegraphics[width=0.45\linewidth]{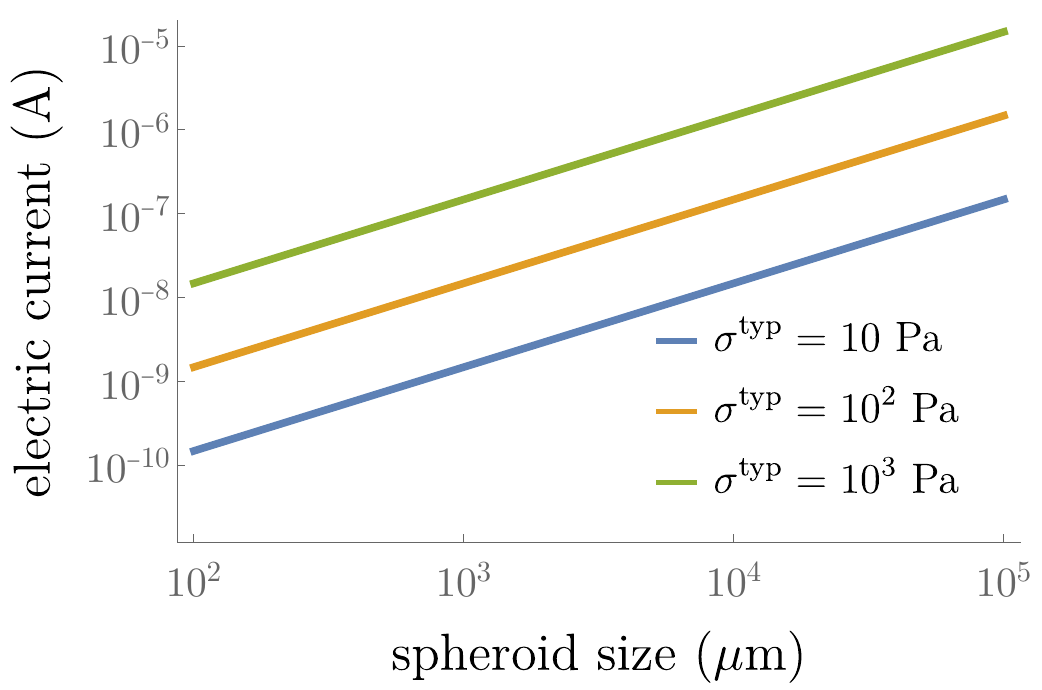}}
	\caption{Orders of magnitude of the external fields required to observe a significant change in the spheroid dynamics as a function of its size as given by Eqs.~\eqref{eq_orderMagnitude_Q}-\eqref{eq_orderMagnitude_I}. \textbf{(a)} Volumetric flow $Q^{\rm ext}$. \textbf{(b)} Electric current $I^{\rm ext}$.
	}
    \label{fig_orderMagnitude}
\end{figure}

\section{Numerical solution to the dynamics equations}
\label{sec_numerics}

Figures~\ref{fig_protocols} and~\ref{fig_protocols_current} of the main text have been obtained by solving the coupled equations~\eqref{eq_r1_Q} and~\eqref{eq_r2_Q} using the dimensionless parameters given in Table~\ref{table_parameters_plots}. The presence of the drain provides a natural cut-off for the (dimensionless) lumen radius $r_1$: if at a time $t_0$ the inner radius $R_1(t)$ reaches the size of the outer drain radius, which we assume to be equal to $2R_{\rm d}=10~\mu$m, we fix its (dimensionless) value to be $r_1(t>t_0)=2R_{\rm d}/R_0$ and solve only Eq.~\eqref{eq_r2_Q} with this fixed value for $r_1$. If the external flow is stopped at a subsequent time $t_1>t_0$ (see Fig.~\ref{fig_protocols_unsucess2} of the main text for instance), we restart solving simultaneously Eqs.~\eqref{eq_r1_Q} and~\eqref{eq_r2_Q}, taking $r_1(t_1)=2R_{\rm d}/R_0$ as the initial condition.

Figure~\ref{fig_stateDiagrams} of the main text has been obtained by taking \mbox{$r_1=2R_{\rm d}/R_0$}  and solving numerically equation~\eqref{eq_r2_Q} using the dimensionless parameters given in Table~\ref{table_parameters_plots}, with the initial condition $R_2(t=0)=R_2^0$ and a given imposed external flux $Q^{\rm ext}$. If a spheroid reaches a cut-off size $R_2=2R_{\rm d}+h_{\rm c}$ after a time $t=t_{\rm s}$, it is classified as ``suppressed'' and its suppression time $t_{\rm s}$ is used to produce the color coding.

\section{Estimates for parameter values}
\label{sec_estimations}

Estimation of the phenomenological parameters appearing in the continuum model is crucial for our analysis. Some of these parameters, such as the cell shear and bulk viscosities, the tissue surface tension and the bulk permeability for instance, have already been estimated in experiments. For most of the remaining parameters, experimental values are not yet available and we have used order-of-magnitude estimations (see Refs.~\cite{sarkar2019field} and~\cite{duclut2019fluid} for details). We provide parameters values and references for these values in Table~\ref{table_estimationParam}.

\begin{table*}[t]
    \setlength{\tabcolsep}{8pt}
    \centering
    \begin{tabular}{lllllllllllllll}
        \hline\hline
        \multirow{2}{*}{\textbf{Figure}} & \multicolumn{14}{c}{\textbf{Parameters values}}\\
        & $\delta_1$ & $\delta_2$ & $\chi$ & $\hat\gamma_0$  & $\hat\gamma_1$ & $\hat\gamma_2$ & $\hat \lambda$ & $\hat V_1$ & $\hat V_2$ & $u_1$ & $u_2$ & $u_3$ & $j_2$ & $j_3$   \\
        \hline
        \ref{fig_protocols} & 0.1 & -1 & 1 & 0 & -0.05 & 0.15 & 0 & 1 & 0.1 & $10^2$ & 0 & 0 & 0 & 0  \\
        \ref{fig_protocols_current} & 0.3 & -1 & 1 & 0 & 0.15 & 0.15 & 0.2 & 0.1 & 0.5 & $10^2$ & 0 & 0 & 0 & 0  \\
        \ref{fig_stateDiagram_1} & - & -1 & 1 & 0 & - & 0.15 & 0.5 & - & 0.3 & $10^2$ & 0 & 0 & 0 & 0  \\
        \ref{fig_stateDiagram_2} & - & 1 & 1 & -0.15 & - & 0.15 & 5.5 & - & 0 & $10^2$ & 0 & 0 & 0 & 0  \\
        \ref{fig_stateDiagram_3} & - & 1 & 1 & 0 & - & 0.15 & 4 & - & 0.15 & $10^2$ & 0 & 0 & 0 & 0  \\
        \hline\hline
    \end{tabular}
    \caption{Dimensionless parameter values used for plotting the figures. These parameters are defined in Eqs.~\eqref{eq_dimless_parameters} to~\eqref{eq_dimLess_all_uj} in App.~\ref{sec_derivation_radii}.
    }
    \label{table_parameters_plots}
\end{table*}

\clearpage

\bibliography{biblio.bib}

\begin{thebibliography}{53}%
\makeatletter
\providecommand \@ifxundefined [1]{%
 \@ifx{#1\undefined}
}%
\providecommand \@ifnum [1]{%
 \ifnum #1\expandafter \@firstoftwo
 \else \expandafter \@secondoftwo
 \fi
}%
\providecommand \@ifx [1]{%
 \ifx #1\expandafter \@firstoftwo
 \else \expandafter \@secondoftwo
 \fi
}%
\providecommand \natexlab [1]{#1}%
\providecommand \enquote  [1]{``#1''}%
\providecommand \bibnamefont  [1]{#1}%
\providecommand \bibfnamefont [1]{#1}%
\providecommand \citenamefont [1]{#1}%
\providecommand \href@noop [0]{\@secondoftwo}%
\providecommand \href [0]{\begingroup \@sanitize@url \@href}%
\providecommand \@href[1]{\@@startlink{#1}\@@href}%
\providecommand \@@href[1]{\endgroup#1\@@endlink}%
\providecommand \@sanitize@url [0]{\catcode `\\12\catcode `\$12\catcode
  `\&12\catcode `\#12\catcode `\^12\catcode `\_12\catcode `\%12\relax}%
\providecommand \@@startlink[1]{}%
\providecommand \@@endlink[0]{}%
\providecommand \url  [0]{\begingroup\@sanitize@url \@url }%
\providecommand \@url [1]{\endgroup\@href {#1}{\urlprefix }}%
\providecommand \urlprefix  [0]{URL }%
\providecommand \Eprint [0]{\href }%
\providecommand \doibase [0]{http://dx.doi.org/}%
\providecommand \selectlanguage [0]{\@gobble}%
\providecommand \bibinfo  [0]{\@secondoftwo}%
\providecommand \bibfield  [0]{\@secondoftwo}%
\providecommand \translation [1]{[#1]}%
\providecommand \BibitemOpen [0]{}%
\providecommand \bibitemStop [0]{}%
\providecommand \bibitemNoStop [0]{.\EOS\space}%
\providecommand \EOS [0]{\spacefactor3000\relax}%
\providecommand \BibitemShut  [1]{\csname bibitem#1\endcsname}%
\let\auto@bib@innerbib\@empty
\bibitem [{\citenamefont {Green}\ and\ \citenamefont
  {Sharpe}(2015)}]{green2015positional}%
  \BibitemOpen
  \bibfield  {author} {\bibinfo {author} {\bibfnamefont {J.~B.~A.}\
  \bibnamefont {Green}}\ and\ \bibinfo {author} {\bibfnamefont
  {J.}~\bibnamefont {Sharpe}},\ }\bibfield  {title} {\enquote {\bibinfo {title}
  {Positional information and reaction-diffusion: Two big ideas in
  developmental biology combine},}\ }\href {\doibase 10.1242/dev.114991}
  {\bibfield  {journal} {\bibinfo  {journal} {Development}\ }\textbf {\bibinfo
  {volume} {142}},\ \bibinfo {pages} {1203--1211} (\bibinfo {year}
  {2015})}\BibitemShut {NoStop}%
\bibitem [{\citenamefont {Miller}\ and\ \citenamefont
  {Davidson}(2013)}]{miller2013interplay}%
  \BibitemOpen
  \bibfield  {author} {\bibinfo {author} {\bibfnamefont {C.~J.}\ \bibnamefont
  {Miller}}\ and\ \bibinfo {author} {\bibfnamefont {L.~A.}\ \bibnamefont
  {Davidson}},\ }\bibfield  {title} {\enquote {\bibinfo {title} {The interplay
  between cell signalling and mechanics in developmental processes},}\ }\href
  {\doibase 10.1038/nrg3513} {\bibfield  {journal} {\bibinfo  {journal} {Nat.
  Rev. Genet.}\ }\textbf {\bibinfo {volume} {14}},\ \bibinfo {pages} {733--744}
  (\bibinfo {year} {2013})}\BibitemShut {NoStop}%
\bibitem [{\citenamefont {Mammoto}\ \emph {et~al.}(2013)\citenamefont
  {Mammoto}, \citenamefont {Mammoto},\ and\ \citenamefont
  {Ingber}}]{mammoto2013mechanobiology}%
  \BibitemOpen
  \bibfield  {author} {\bibinfo {author} {\bibfnamefont {T.}~\bibnamefont
  {Mammoto}}, \bibinfo {author} {\bibfnamefont {A.}~\bibnamefont {Mammoto}}, \
  and\ \bibinfo {author} {\bibfnamefont {D.~E.}\ \bibnamefont {Ingber}},\
  }\bibfield  {title} {\enquote {\bibinfo {title} {Mechanobiology and
  {{Developmental Control}}},}\ }\href {\doibase
  10.1146/annurev-cellbio-101512-122340} {\bibfield  {journal} {\bibinfo
  {journal} {Annu. Rev. Cell Dev. Biol.}\ }\textbf {\bibinfo {volume} {29}},\
  \bibinfo {pages} {27--61} (\bibinfo {year} {2013})}\BibitemShut {NoStop}%
\bibitem [{\citenamefont {Ladoux}\ and\ \citenamefont
  {M{\`e}ge}(2017)}]{ladoux2017mechanobiology}%
  \BibitemOpen
  \bibfield  {author} {\bibinfo {author} {\bibfnamefont {B.}~\bibnamefont
  {Ladoux}}\ and\ \bibinfo {author} {\bibfnamefont {R.-M.}\ \bibnamefont
  {M{\`e}ge}},\ }\bibfield  {title} {\enquote {\bibinfo {title} {Mechanobiology
  of collective cell behaviours},}\ }\href {\doibase 10.1038/nrm.2017.98}
  {\bibfield  {journal} {\bibinfo  {journal} {Nat. Rev. Mol. Cell Biol.}\
  }\textbf {\bibinfo {volume} {18}},\ \bibinfo {pages} {743--757} (\bibinfo
  {year} {2017})}\BibitemShut {NoStop}%
\bibitem [{\citenamefont {{Ruiz-Herrero}}\ \emph {et~al.}(2017)\citenamefont
  {{Ruiz-Herrero}}, \citenamefont {Alessandri}, \citenamefont {Gurchenkov},
  \citenamefont {Nassoy},\ and\ \citenamefont
  {Mahadevan}}]{ruiz-herrero2017organ}%
  \BibitemOpen
  \bibfield  {author} {\bibinfo {author} {\bibfnamefont {T.}~\bibnamefont
  {{Ruiz-Herrero}}}, \bibinfo {author} {\bibfnamefont {K.}~\bibnamefont
  {Alessandri}}, \bibinfo {author} {\bibfnamefont {B.~V.}\ \bibnamefont
  {Gurchenkov}}, \bibinfo {author} {\bibfnamefont {P.}~\bibnamefont {Nassoy}},
  \ and\ \bibinfo {author} {\bibfnamefont {L.}~\bibnamefont {Mahadevan}},\
  }\bibfield  {title} {\enquote {\bibinfo {title} {Organ size control via
  hydraulically gated oscillations},}\ }\href {\doibase 10.1242/dev.153056}
  {\bibfield  {journal} {\bibinfo  {journal} {Development}\ }\textbf {\bibinfo
  {volume} {144}},\ \bibinfo {pages} {4422--4427} (\bibinfo {year}
  {2017})}\BibitemShut {NoStop}%
\bibitem [{\citenamefont {Dumortier}\ \emph {et~al.}(2019)\citenamefont
  {Dumortier}, \citenamefont {{Le Verge-Serandour}}, \citenamefont
  {Tortorelli}, \citenamefont {Mielke}, \citenamefont {{de Plater}},
  \citenamefont {Turlier},\ and\ \citenamefont
  {Ma{\^i}tre}}]{dumortier2019hydraulic}%
  \BibitemOpen
  \bibfield  {author} {\bibinfo {author} {\bibfnamefont {J.~G.}\ \bibnamefont
  {Dumortier}}, \bibinfo {author} {\bibfnamefont {M.}~\bibnamefont {{Le
  Verge-Serandour}}}, \bibinfo {author} {\bibfnamefont {A.~F.}\ \bibnamefont
  {Tortorelli}}, \bibinfo {author} {\bibfnamefont {A.}~\bibnamefont {Mielke}},
  \bibinfo {author} {\bibfnamefont {L.}~\bibnamefont {{de Plater}}}, \bibinfo
  {author} {\bibfnamefont {H.}~\bibnamefont {Turlier}}, \ and\ \bibinfo
  {author} {\bibfnamefont {J.-L.}\ \bibnamefont {Ma{\^i}tre}},\ }\bibfield
  {title} {\enquote {\bibinfo {title} {Hydraulic fracturing and active
  coarsening position the lumen of the mouse blastocyst},}\ }\href {\doibase
  10.1126/science.aaw7709} {\bibfield  {journal} {\bibinfo  {journal}
  {Science}\ }\textbf {\bibinfo {volume} {365}},\ \bibinfo {pages} {465--468}
  (\bibinfo {year} {2019})}\BibitemShut {NoStop}%
\bibitem [{\citenamefont {Levin}\ and\ \citenamefont
  {Martyniuk}(2018)}]{levin2018bioelectric}%
  \BibitemOpen
  \bibfield  {author} {\bibinfo {author} {\bibfnamefont {M.}~\bibnamefont
  {Levin}}\ and\ \bibinfo {author} {\bibfnamefont {C.~J.}\ \bibnamefont
  {Martyniuk}},\ }\bibfield  {title} {\enquote {\bibinfo {title} {The
  bioelectric code: {{An}} ancient computational medium for dynamic control of
  growth and form},}\ }\href {\doibase 10.1016/j.biosystems.2017.08.009}
  {\bibfield  {journal} {\bibinfo  {journal} {Biosystems}\ }\textbf {\bibinfo
  {volume} {164}},\ \bibinfo {pages} {76--93} (\bibinfo {year}
  {2018})}\BibitemShut {NoStop}%
\bibitem [{\citenamefont {Silver}\ and\ \citenamefont
  {Nelson}(2018)}]{silver2018bioelectric}%
  \BibitemOpen
  \bibfield  {author} {\bibinfo {author} {\bibfnamefont {B.~B.}\ \bibnamefont
  {Silver}}\ and\ \bibinfo {author} {\bibfnamefont {C.~M.}\ \bibnamefont
  {Nelson}},\ }\bibfield  {title} {\enquote {\bibinfo {title} {The
  {{Bioelectric Code}}: {{Reprogramming Cancer}} and {{Aging From}} the
  {{Interface}} of {{Mechanical}} and {{Chemical Microenvironments}}},}\ }\href
  {\doibase 10.3389/fcell.2018.00021} {\bibfield  {journal} {\bibinfo
  {journal} {Front. Cell Dev. Biol.}\ }\textbf {\bibinfo {volume} {6}},\
  \bibinfo {pages} {21} (\bibinfo {year} {2018})}\BibitemShut {NoStop}%
\bibitem [{\citenamefont {Chan}\ \emph {et~al.}(2019)\citenamefont {Chan},
  \citenamefont {Costanzo}, \citenamefont {{Ruiz-Herrero}}, \citenamefont
  {M{\"o}nke}, \citenamefont {Petrie}, \citenamefont {Bergert}, \citenamefont
  {{Diz-Mu{\~n}oz}}, \citenamefont {Mahadevan},\ and\ \citenamefont
  {Hiiragi}}]{chan2019hydraulic}%
  \BibitemOpen
  \bibfield  {author} {\bibinfo {author} {\bibfnamefont {C.~J.}\ \bibnamefont
  {Chan}}, \bibinfo {author} {\bibfnamefont {M.}~\bibnamefont {Costanzo}},
  \bibinfo {author} {\bibfnamefont {T.}~\bibnamefont {{Ruiz-Herrero}}},
  \bibinfo {author} {\bibfnamefont {G.}~\bibnamefont {M{\"o}nke}}, \bibinfo
  {author} {\bibfnamefont {R.~J.}\ \bibnamefont {Petrie}}, \bibinfo {author}
  {\bibfnamefont {M.}~\bibnamefont {Bergert}}, \bibinfo {author} {\bibfnamefont
  {A.}~\bibnamefont {{Diz-Mu{\~n}oz}}}, \bibinfo {author} {\bibfnamefont
  {L.}~\bibnamefont {Mahadevan}}, \ and\ \bibinfo {author} {\bibfnamefont
  {T.}~\bibnamefont {Hiiragi}},\ }\bibfield  {title} {\enquote {\bibinfo
  {title} {Hydraulic control of mammalian embryo size and cell fate},}\ }\href
  {\doibase 10.1038/s41586-019-1309-x} {\bibfield  {journal} {\bibinfo
  {journal} {Nature}\ }\textbf {\bibinfo {volume} {571}},\ \bibinfo {pages}
  {112--116} (\bibinfo {year} {2019})}\BibitemShut {NoStop}%
\bibitem [{\citenamefont {F{\"u}tterer}\ \emph {et~al.}(2003)\citenamefont
  {F{\"u}tterer}, \citenamefont {Colombo}, \citenamefont {J{\"u}licher},\ and\
  \citenamefont {Ott}}]{futterer2003morphogenetic}%
  \BibitemOpen
  \bibfield  {author} {\bibinfo {author} {\bibfnamefont {C.}~\bibnamefont
  {F{\"u}tterer}}, \bibinfo {author} {\bibfnamefont {C.}~\bibnamefont
  {Colombo}}, \bibinfo {author} {\bibfnamefont {F.}~\bibnamefont
  {J{\"u}licher}}, \ and\ \bibinfo {author} {\bibfnamefont {A.}~\bibnamefont
  {Ott}},\ }\bibfield  {title} {\enquote {\bibinfo {title} {Morphogenetic
  oscillations during symmetry breaking of regenerating
  {{{\emph{Hydra}}}}{\emph{ vulgaris}} cells},}\ }\href {\doibase
  10.1209/epl/i2003-00148-y} {\bibfield  {journal} {\bibinfo  {journal} {EPL}\
  }\textbf {\bibinfo {volume} {64}},\ \bibinfo {pages} {137--143} (\bibinfo
  {year} {2003})}\BibitemShut {NoStop}%
\bibitem [{\citenamefont {Roux}(1892)}]{roux1892uber}%
  \BibitemOpen
  \bibfield  {author} {\bibinfo {author} {\bibfnamefont {W.}~\bibnamefont
  {Roux}},\ }\bibfield  {title} {\enquote {\bibinfo {title} {\"uber die
  morphologische {{Polarisation}} von {{Eiern}} und {{Embryonen}} durch den
  electrischen {{Strom}}},}\ }\href@noop {} {\bibfield  {journal} {\bibinfo
  {journal} {Math. Naturwiss. Kl.}\ }\textbf {\bibinfo {volume} {101}},\
  \bibinfo {pages} {27--228} (\bibinfo {year} {1892})}\BibitemShut {NoStop}%
\bibitem [{\citenamefont {McCaig}\ \emph {et~al.}(2005)\citenamefont {McCaig},
  \citenamefont {Rajnicek}, \citenamefont {Song},\ and\ \citenamefont
  {Zhao}}]{mccaig2005controlling}%
  \BibitemOpen
  \bibfield  {author} {\bibinfo {author} {\bibfnamefont {C.~D.}\ \bibnamefont
  {McCaig}}, \bibinfo {author} {\bibfnamefont {A.~M.}\ \bibnamefont
  {Rajnicek}}, \bibinfo {author} {\bibfnamefont {B.}~\bibnamefont {Song}}, \
  and\ \bibinfo {author} {\bibfnamefont {M.}~\bibnamefont {Zhao}},\ }\bibfield
  {title} {\enquote {\bibinfo {title} {Controlling {{Cell Behavior
  Electrically}}: {{Current Views}} and {{Future Potential}}},}\ }\href
  {\doibase 10.1152/physrev.00020.2004} {\bibfield  {journal} {\bibinfo
  {journal} {Physiol. Rev.}\ }\textbf {\bibinfo {volume} {85}},\ \bibinfo
  {pages} {943--978} (\bibinfo {year} {2005})}\BibitemShut {NoStop}%
\bibitem [{\citenamefont {McLaughlin}\ and\ \citenamefont
  {Levin}(2018)}]{mclaughlin2018bioelectric}%
  \BibitemOpen
  \bibfield  {author} {\bibinfo {author} {\bibfnamefont {K.~A.}\ \bibnamefont
  {McLaughlin}}\ and\ \bibinfo {author} {\bibfnamefont {M.}~\bibnamefont
  {Levin}},\ }\bibfield  {title} {\enquote {\bibinfo {title} {Bioelectric
  signaling in regeneration: {{Mechanisms}} of ionic controls of growth and
  form},}\ }\href {\doibase 10.1016/j.ydbio.2017.08.032} {\bibfield  {journal}
  {\bibinfo  {journal} {Dev. Biol.}\ }\textbf {\bibinfo {volume} {433}},\
  \bibinfo {pages} {177--189} (\bibinfo {year} {2018})}\BibitemShut {NoStop}%
\bibitem [{\citenamefont {Watanabe}\ \emph {et~al.}(2009)\citenamefont
  {Watanabe}, \citenamefont {Fujisawa},\ and\ \citenamefont
  {Holstein}}]{watanabe2009cnidarians}%
  \BibitemOpen
  \bibfield  {author} {\bibinfo {author} {\bibfnamefont {H.}~\bibnamefont
  {Watanabe}}, \bibinfo {author} {\bibfnamefont {T.}~\bibnamefont {Fujisawa}},
  \ and\ \bibinfo {author} {\bibfnamefont {T.~W.}\ \bibnamefont {Holstein}},\
  }\bibfield  {title} {\enquote {\bibinfo {title} {Cnidarians and the
  evolutionary origin of the nervous system: {{Cnidarian}} nervous system},}\
  }\href {\doibase 10.1111/j.1440-169X.2009.01103.x} {\bibfield  {journal}
  {\bibinfo  {journal} {Develop. Growth Differ.}\ }\textbf {\bibinfo {volume}
  {51}},\ \bibinfo {pages} {167--183} (\bibinfo {year} {2009})}\BibitemShut
  {NoStop}%
\bibitem [{\citenamefont {Braun}\ and\ \citenamefont
  {Ori}(2019)}]{braun2019electricinduced}%
  \BibitemOpen
  \bibfield  {author} {\bibinfo {author} {\bibfnamefont {E.}~\bibnamefont
  {Braun}}\ and\ \bibinfo {author} {\bibfnamefont {H.}~\bibnamefont {Ori}},\
  }\bibfield  {title} {\enquote {\bibinfo {title} {Electric-{{Induced
  Reversal}} of {{Morphogenesis}} in {{Hydra}}},}\ }\href {\doibase
  10.1016/j.bpj.2019.09.007} {\bibfield  {journal} {\bibinfo  {journal}
  {Biophys. J.}\ }\textbf {\bibinfo {volume} {117}},\ \bibinfo {pages}
  {1514--1523} (\bibinfo {year} {2019})}\BibitemShut {NoStop}%
\bibitem [{\citenamefont {{Mart{\'i}n-Belmonte}}\ \emph
  {et~al.}(2008)\citenamefont {{Mart{\'i}n-Belmonte}}, \citenamefont {Yu},
  \citenamefont {{Rodr{\'i}guez-Fraticelli}}, \citenamefont {Ewald},
  \citenamefont {Werb}, \citenamefont {Alonso},\ and\ \citenamefont
  {Mostov}}]{martin-belmonte2008cellpolarity}%
  \BibitemOpen
  \bibfield  {author} {\bibinfo {author} {\bibfnamefont {F.}~\bibnamefont
  {{Mart{\'i}n-Belmonte}}}, \bibinfo {author} {\bibfnamefont {W.}~\bibnamefont
  {Yu}}, \bibinfo {author} {\bibfnamefont {A.}~\bibnamefont
  {{Rodr{\'i}guez-Fraticelli}}}, \bibinfo {author} {\bibfnamefont
  {A.}~\bibnamefont {Ewald}}, \bibinfo {author} {\bibfnamefont
  {Z.}~\bibnamefont {Werb}}, \bibinfo {author} {\bibfnamefont {M.~A.}\
  \bibnamefont {Alonso}}, \ and\ \bibinfo {author} {\bibfnamefont
  {K.}~\bibnamefont {Mostov}},\ }\bibfield  {title} {\enquote {\bibinfo {title}
  {Cell-{{Polarity Dynamics Controls}} the {{Mechanism}} of {{Lumen Formation}}
  in {{Epithelial Morphogenesis}}},}\ }\href {\doibase
  10.1016/j.cub.2008.02.076} {\bibfield  {journal} {\bibinfo  {journal} {Curr.
  Biol.}\ }\textbf {\bibinfo {volume} {18}},\ \bibinfo {pages} {507--513}
  (\bibinfo {year} {2008})}\BibitemShut {NoStop}%
\bibitem [{\citenamefont {Debnath}\ \emph {et~al.}(2002)\citenamefont
  {Debnath}, \citenamefont {Mills}, \citenamefont {Collins}, \citenamefont
  {Reginato}, \citenamefont {Muthuswamy},\ and\ \citenamefont
  {Brugge}}]{debnath2002role}%
  \BibitemOpen
  \bibfield  {author} {\bibinfo {author} {\bibfnamefont {J.}~\bibnamefont
  {Debnath}}, \bibinfo {author} {\bibfnamefont {K.~R.}\ \bibnamefont {Mills}},
  \bibinfo {author} {\bibfnamefont {N.~L.}\ \bibnamefont {Collins}}, \bibinfo
  {author} {\bibfnamefont {M.~J.}\ \bibnamefont {Reginato}}, \bibinfo {author}
  {\bibfnamefont {S.~K.}\ \bibnamefont {Muthuswamy}}, \ and\ \bibinfo {author}
  {\bibfnamefont {J.~S.}\ \bibnamefont {Brugge}},\ }\bibfield  {title}
  {\enquote {\bibinfo {title} {The {{Role}} of {{Apoptosis}} in {{Creating}}
  and {{Maintaining Luminal Space}} within {{Normal}} and
  {{Oncogene}}-{{Expressing Mammary Acini}}},}\ }\href {\doibase
  10.1016/S0092-8674(02)01001-2} {\bibfield  {journal} {\bibinfo  {journal}
  {Cell}\ }\textbf {\bibinfo {volume} {111}},\ \bibinfo {pages} {29--40}
  (\bibinfo {year} {2002})}\BibitemShut {NoStop}%
\bibitem [{\citenamefont {Huch}\ and\ \citenamefont
  {Koo}(2015)}]{huch2015modeling}%
  \BibitemOpen
  \bibfield  {author} {\bibinfo {author} {\bibfnamefont {M.}~\bibnamefont
  {Huch}}\ and\ \bibinfo {author} {\bibfnamefont {B.-K.}\ \bibnamefont {Koo}},\
  }\bibfield  {title} {\enquote {\bibinfo {title} {Modeling mouse and human
  development using organoid cultures},}\ }\href {\doibase 10.1242/dev.118570}
  {\bibfield  {journal} {\bibinfo  {journal} {Development}\ }\textbf {\bibinfo
  {volume} {142}},\ \bibinfo {pages} {3113--3125} (\bibinfo {year}
  {2015})}\BibitemShut {NoStop}%
\bibitem [{\citenamefont {{Dahl-Jensen}}\ and\ \citenamefont
  {{Grapin-Botton}}(2017)}]{dahl-jensen2017physics}%
  \BibitemOpen
  \bibfield  {author} {\bibinfo {author} {\bibfnamefont {S.}~\bibnamefont
  {{Dahl-Jensen}}}\ and\ \bibinfo {author} {\bibfnamefont {A.}~\bibnamefont
  {{Grapin-Botton}}},\ }\bibfield  {title} {\enquote {\bibinfo {title} {The
  physics of organoids: A biophysical approach to understanding
  organogenesis},}\ }\href {\doibase 10.1242/dev.143693} {\bibfield  {journal}
  {\bibinfo  {journal} {Development}\ }\textbf {\bibinfo {volume} {144}},\
  \bibinfo {pages} {946--951} (\bibinfo {year} {2017})}\BibitemShut {NoStop}%
\bibitem [{\citenamefont {Alcaraz}\ \emph {et~al.}(2008)\citenamefont
  {Alcaraz}, \citenamefont {Xu}, \citenamefont {Mori}, \citenamefont {Nelson},
  \citenamefont {Mroue}, \citenamefont {Spencer}, \citenamefont {Brownfield},
  \citenamefont {Radisky}, \citenamefont {Bustamante},\ and\ \citenamefont
  {Bissell}}]{alcaraz2008laminin}%
  \BibitemOpen
  \bibfield  {author} {\bibinfo {author} {\bibfnamefont {J.}~\bibnamefont
  {Alcaraz}}, \bibinfo {author} {\bibfnamefont {R.}~\bibnamefont {Xu}},
  \bibinfo {author} {\bibfnamefont {H.}~\bibnamefont {Mori}}, \bibinfo {author}
  {\bibfnamefont {C.~M.}\ \bibnamefont {Nelson}}, \bibinfo {author}
  {\bibfnamefont {R.}~\bibnamefont {Mroue}}, \bibinfo {author} {\bibfnamefont
  {V.~A.}\ \bibnamefont {Spencer}}, \bibinfo {author} {\bibfnamefont
  {D.}~\bibnamefont {Brownfield}}, \bibinfo {author} {\bibfnamefont {D.~C.}\
  \bibnamefont {Radisky}}, \bibinfo {author} {\bibfnamefont {C.}~\bibnamefont
  {Bustamante}}, \ and\ \bibinfo {author} {\bibfnamefont {M.~J.}\ \bibnamefont
  {Bissell}},\ }\bibfield  {title} {\enquote {\bibinfo {title} {Laminin and
  biomimetic extracellular elasticity enhance functional differentiation in
  mammary epithelia},}\ }\href {\doibase 10.1038/emboj.2008.206} {\bibfield
  {journal} {\bibinfo  {journal} {EMBO J.}\ }\textbf {\bibinfo {volume} {27}},\
  \bibinfo {pages} {2829--2838} (\bibinfo {year} {2008})}\BibitemShut {NoStop}%
\bibitem [{\citenamefont {Montel}\ \emph {et~al.}(2011)\citenamefont {Montel},
  \citenamefont {Delarue}, \citenamefont {Elgeti}, \citenamefont {Malaquin},
  \citenamefont {Basan}, \citenamefont {Risler}, \citenamefont {Cabane},
  \citenamefont {Vignjevic}, \citenamefont {Prost}, \citenamefont {Cappello},\
  and\ \citenamefont {Joanny}}]{montel2011stress}%
  \BibitemOpen
  \bibfield  {author} {\bibinfo {author} {\bibfnamefont {F.}~\bibnamefont
  {Montel}}, \bibinfo {author} {\bibfnamefont {M.}~\bibnamefont {Delarue}},
  \bibinfo {author} {\bibfnamefont {J.}~\bibnamefont {Elgeti}}, \bibinfo
  {author} {\bibfnamefont {L.}~\bibnamefont {Malaquin}}, \bibinfo {author}
  {\bibfnamefont {M.}~\bibnamefont {Basan}}, \bibinfo {author} {\bibfnamefont
  {T.}~\bibnamefont {Risler}}, \bibinfo {author} {\bibfnamefont
  {B.}~\bibnamefont {Cabane}}, \bibinfo {author} {\bibfnamefont
  {D.}~\bibnamefont {Vignjevic}}, \bibinfo {author} {\bibfnamefont
  {J.}~\bibnamefont {Prost}}, \bibinfo {author} {\bibfnamefont
  {G.}~\bibnamefont {Cappello}}, \ and\ \bibinfo {author} {\bibfnamefont
  {J.-F.}\ \bibnamefont {Joanny}},\ }\bibfield  {title} {\enquote {\bibinfo
  {title} {Stress {{Clamp Experiments}} on {{Multicellular Tumor
  Spheroids}}},}\ }\href {\doibase 10.1103/PhysRevLett.107.188102} {\bibfield
  {journal} {\bibinfo  {journal} {Phys. Rev. Lett.}\ }\textbf {\bibinfo
  {volume} {107}},\ \bibinfo {pages} {188102} (\bibinfo {year}
  {2011})}\BibitemShut {NoStop}%
\bibitem [{\citenamefont {Delarue}\ \emph {et~al.}(2013)\citenamefont
  {Delarue}, \citenamefont {Montel}, \citenamefont {Caen}, \citenamefont
  {Elgeti}, \citenamefont {Siaugue}, \citenamefont {Vignjevic}, \citenamefont
  {Prost}, \citenamefont {Joanny},\ and\ \citenamefont
  {Cappello}}]{delarue2013mechanical}%
  \BibitemOpen
  \bibfield  {author} {\bibinfo {author} {\bibfnamefont {M.}~\bibnamefont
  {Delarue}}, \bibinfo {author} {\bibfnamefont {F.}~\bibnamefont {Montel}},
  \bibinfo {author} {\bibfnamefont {O.}~\bibnamefont {Caen}}, \bibinfo {author}
  {\bibfnamefont {J.}~\bibnamefont {Elgeti}}, \bibinfo {author} {\bibfnamefont
  {J.-M.}\ \bibnamefont {Siaugue}}, \bibinfo {author} {\bibfnamefont
  {D.}~\bibnamefont {Vignjevic}}, \bibinfo {author} {\bibfnamefont
  {J.}~\bibnamefont {Prost}}, \bibinfo {author} {\bibfnamefont {J.-F.}\
  \bibnamefont {Joanny}}, \ and\ \bibinfo {author} {\bibfnamefont
  {G.}~\bibnamefont {Cappello}},\ }\bibfield  {title} {\enquote {\bibinfo
  {title} {Mechanical {{Control}} of {{Cell}} flow in {{Multicellular
  Spheroids}}},}\ }\href {\doibase 10.1103/PhysRevLett.110.138103} {\bibfield
  {journal} {\bibinfo  {journal} {Phys. Rev. Lett.}\ }\textbf {\bibinfo
  {volume} {110}},\ \bibinfo {pages} {138103} (\bibinfo {year}
  {2013})}\BibitemShut {NoStop}%
\bibitem [{\citenamefont {Kruse}\ \emph {et~al.}(2005)\citenamefont {Kruse},
  \citenamefont {Joanny}, \citenamefont {J{\"u}licher}, \citenamefont {Prost},\
  and\ \citenamefont {Sekimoto}}]{kruse2005generic}%
  \BibitemOpen
  \bibfield  {author} {\bibinfo {author} {\bibfnamefont {K.}~\bibnamefont
  {Kruse}}, \bibinfo {author} {\bibfnamefont {J.~F.}\ \bibnamefont {Joanny}},
  \bibinfo {author} {\bibfnamefont {F.}~\bibnamefont {J{\"u}licher}}, \bibinfo
  {author} {\bibfnamefont {J.}~\bibnamefont {Prost}}, \ and\ \bibinfo {author}
  {\bibfnamefont {K.}~\bibnamefont {Sekimoto}},\ }\bibfield  {title} {\enquote
  {\bibinfo {title} {Generic theory of active polar gels: A paradigm for
  cytoskeletal dynamics},}\ }\href {\doibase 10.1140/epje/e2005-00002-5}
  {\bibfield  {journal} {\bibinfo  {journal} {Eur. Phys. J. E}\ }\textbf
  {\bibinfo {volume} {16}},\ \bibinfo {pages} {5--16} (\bibinfo {year}
  {2005})}\BibitemShut {NoStop}%
\bibitem [{\citenamefont {Marchetti}\ \emph {et~al.}(2013)\citenamefont
  {Marchetti}, \citenamefont {Joanny}, \citenamefont {Ramaswamy}, \citenamefont
  {Liverpool}, \citenamefont {Prost}, \citenamefont {Rao},\ and\ \citenamefont
  {Simha}}]{marchetti2013hydrodynamics}%
  \BibitemOpen
  \bibfield  {author} {\bibinfo {author} {\bibfnamefont {M.~C.}\ \bibnamefont
  {Marchetti}}, \bibinfo {author} {\bibfnamefont {J.~F.}\ \bibnamefont
  {Joanny}}, \bibinfo {author} {\bibfnamefont {S.}~\bibnamefont {Ramaswamy}},
  \bibinfo {author} {\bibfnamefont {T.~B.}\ \bibnamefont {Liverpool}}, \bibinfo
  {author} {\bibfnamefont {J.}~\bibnamefont {Prost}}, \bibinfo {author}
  {\bibfnamefont {M.}~\bibnamefont {Rao}}, \ and\ \bibinfo {author}
  {\bibfnamefont {R.~A.}\ \bibnamefont {Simha}},\ }\bibfield  {title} {\enquote
  {\bibinfo {title} {Hydrodynamics of soft active matter},}\ }\href {\doibase
  10.1103/RevModPhys.85.1143} {\bibfield  {journal} {\bibinfo  {journal} {Rev.
  Mod. Phys.}\ }\textbf {\bibinfo {volume} {85}},\ \bibinfo {pages}
  {1143--1189} (\bibinfo {year} {2013})}\BibitemShut {NoStop}%
\bibitem [{\citenamefont {Ranft}\ \emph {et~al.}(2010)\citenamefont {Ranft},
  \citenamefont {Basan}, \citenamefont {Elgeti}, \citenamefont {Joanny},
  \citenamefont {Prost},\ and\ \citenamefont
  {J{\"u}licher}}]{ranft2010fluidization}%
  \BibitemOpen
  \bibfield  {author} {\bibinfo {author} {\bibfnamefont {J.}~\bibnamefont
  {Ranft}}, \bibinfo {author} {\bibfnamefont {M.}~\bibnamefont {Basan}},
  \bibinfo {author} {\bibfnamefont {J.}~\bibnamefont {Elgeti}}, \bibinfo
  {author} {\bibfnamefont {J.-F.}\ \bibnamefont {Joanny}}, \bibinfo {author}
  {\bibfnamefont {J.}~\bibnamefont {Prost}}, \ and\ \bibinfo {author}
  {\bibfnamefont {F.}~\bibnamefont {J{\"u}licher}},\ }\bibfield  {title}
  {\enquote {\bibinfo {title} {Fluidization of tissues by cell division and
  apoptosis},}\ }\href {\doibase 10.1073/pnas.1011086107} {\bibfield  {journal}
  {\bibinfo  {journal} {Proc. Natl. Acad. Sci. U.S.A.}\ }\textbf {\bibinfo
  {volume} {107}},\ \bibinfo {pages} {20863--20868} (\bibinfo {year}
  {2010})}\BibitemShut {NoStop}%
\bibitem [{\citenamefont {Sarkar}\ \emph {et~al.}(2019)\citenamefont {Sarkar},
  \citenamefont {Prost},\ and\ \citenamefont {J{\"u}licher}}]{sarkar2019field}%
  \BibitemOpen
  \bibfield  {author} {\bibinfo {author} {\bibfnamefont {N.}~\bibnamefont
  {Sarkar}}, \bibinfo {author} {\bibfnamefont {J.}~\bibnamefont {Prost}}, \
  and\ \bibinfo {author} {\bibfnamefont {F.}~\bibnamefont {J{\"u}licher}},\
  }\bibfield  {title} {\enquote {\bibinfo {title} {Field induced cell
  proliferation and death in a model epithelium},}\ }\href {\doibase
  10.1088/1367-2630/ab0a8d} {\bibfield  {journal} {\bibinfo  {journal} {New J.
  Phys.}\ }\textbf {\bibinfo {volume} {21}},\ \bibinfo {pages} {043035}
  (\bibinfo {year} {2019})}\BibitemShut {NoStop}%
\bibitem [{\citenamefont {Duclut}\ \emph {et~al.}(2019)\citenamefont {Duclut},
  \citenamefont {Sarkar}, \citenamefont {Prost},\ and\ \citenamefont
  {J{\"u}licher}}]{duclut2019fluid}%
  \BibitemOpen
  \bibfield  {author} {\bibinfo {author} {\bibfnamefont {C.}~\bibnamefont
  {Duclut}}, \bibinfo {author} {\bibfnamefont {N.}~\bibnamefont {Sarkar}},
  \bibinfo {author} {\bibfnamefont {J.}~\bibnamefont {Prost}}, \ and\ \bibinfo
  {author} {\bibfnamefont {F.}~\bibnamefont {J{\"u}licher}},\ }\bibfield
  {title} {\enquote {\bibinfo {title} {Fluid pumping and active
  flexoelectricity can promote lumen nucleation in cell assemblies},}\ }\href
  {\doibase 10.1073/pnas.1908481116} {\bibfield  {journal} {\bibinfo  {journal}
  {Proc. Natl. Acad. Sci. U.S.A.}\ }\textbf {\bibinfo {volume} {116}},\
  \bibinfo {pages} {19264--19273} (\bibinfo {year} {2019})}\BibitemShut
  {NoStop}%
\bibitem [{\citenamefont {Scott}\ \emph {et~al.}(2012)\citenamefont {Scott},
  \citenamefont {Wolchok},\ and\ \citenamefont {Old}}]{scott2012antibody}%
  \BibitemOpen
  \bibfield  {author} {\bibinfo {author} {\bibfnamefont {A.~M.}\ \bibnamefont
  {Scott}}, \bibinfo {author} {\bibfnamefont {J.~D.}\ \bibnamefont {Wolchok}},
  \ and\ \bibinfo {author} {\bibfnamefont {L.~J.}\ \bibnamefont {Old}},\
  }\bibfield  {title} {\enquote {\bibinfo {title} {Antibody therapy of
  cancer},}\ }\href {\doibase 10.1038/nrc3236} {\bibfield  {journal} {\bibinfo
  {journal} {Nat. Rev. Cancer}\ }\textbf {\bibinfo {volume} {12}},\ \bibinfo
  {pages} {278--287} (\bibinfo {year} {2012})}\BibitemShut {NoStop}%
\bibitem [{\citenamefont {Baskar}\ \emph {et~al.}(2012)\citenamefont {Baskar},
  \citenamefont {Lee}, \citenamefont {Yeo},\ and\ \citenamefont
  {Yeoh}}]{baskar2012cancer}%
  \BibitemOpen
  \bibfield  {author} {\bibinfo {author} {\bibfnamefont {R.}~\bibnamefont
  {Baskar}}, \bibinfo {author} {\bibfnamefont {K.~A.}\ \bibnamefont {Lee}},
  \bibinfo {author} {\bibfnamefont {R.}~\bibnamefont {Yeo}}, \ and\ \bibinfo
  {author} {\bibfnamefont {K.-W.}\ \bibnamefont {Yeoh}},\ }\bibfield  {title}
  {\enquote {\bibinfo {title} {Cancer and {{Radiation Therapy}}: {{Current
  Advances}} and {{Future Directions}}},}\ }\href {\doibase 10.7150/ijms.3635}
  {\bibfield  {journal} {\bibinfo  {journal} {Int. J. Med. Sci.}\ }\textbf
  {\bibinfo {volume} {9}},\ \bibinfo {pages} {193--199} (\bibinfo {year}
  {2012})}\BibitemShut {NoStop}%
\bibitem [{\citenamefont {{Al-Bataineh}}\ \emph {et~al.}(2012)\citenamefont
  {{Al-Bataineh}}, \citenamefont {Jenne},\ and\ \citenamefont
  {Huber}}]{al-bataineh2012clinical}%
  \BibitemOpen
  \bibfield  {author} {\bibinfo {author} {\bibfnamefont {O.}~\bibnamefont
  {{Al-Bataineh}}}, \bibinfo {author} {\bibfnamefont {J.}~\bibnamefont
  {Jenne}}, \ and\ \bibinfo {author} {\bibfnamefont {P.}~\bibnamefont
  {Huber}},\ }\bibfield  {title} {\enquote {\bibinfo {title} {Clinical and
  future applications of high intensity focused ultrasound in cancer},}\ }\href
  {\doibase 10.1016/j.ctrv.2011.08.004} {\bibfield  {journal} {\bibinfo
  {journal} {Cancer Treat. Rev.}\ }\textbf {\bibinfo {volume} {38}},\ \bibinfo
  {pages} {346--353} (\bibinfo {year} {2012})}\BibitemShut {NoStop}%
\bibitem [{\citenamefont {Mittelstein}\ \emph {et~al.}(2020)\citenamefont
  {Mittelstein}, \citenamefont {Ye}, \citenamefont {Schibber}, \citenamefont
  {Roychoudhury}, \citenamefont {Martinez}, \citenamefont {Fekrazad},
  \citenamefont {Ortiz}, \citenamefont {Lee}, \citenamefont {Shapiro},\ and\
  \citenamefont {Gharib}}]{mittelstein2020selective}%
  \BibitemOpen
  \bibfield  {author} {\bibinfo {author} {\bibfnamefont {D.~R.}\ \bibnamefont
  {Mittelstein}}, \bibinfo {author} {\bibfnamefont {J.}~\bibnamefont {Ye}},
  \bibinfo {author} {\bibfnamefont {E.~F.}\ \bibnamefont {Schibber}}, \bibinfo
  {author} {\bibfnamefont {A.}~\bibnamefont {Roychoudhury}}, \bibinfo {author}
  {\bibfnamefont {L.~T.}\ \bibnamefont {Martinez}}, \bibinfo {author}
  {\bibfnamefont {M.~H.}\ \bibnamefont {Fekrazad}}, \bibinfo {author}
  {\bibfnamefont {M.}~\bibnamefont {Ortiz}}, \bibinfo {author} {\bibfnamefont
  {P.~P.}\ \bibnamefont {Lee}}, \bibinfo {author} {\bibfnamefont {M.~G.}\
  \bibnamefont {Shapiro}}, \ and\ \bibinfo {author} {\bibfnamefont
  {M.}~\bibnamefont {Gharib}},\ }\bibfield  {title} {\enquote {\bibinfo {title}
  {Selective ablation of cancer cells with low intensity pulsed ultrasound},}\
  }\href {\doibase 10.1063/1.5128627} {\bibfield  {journal} {\bibinfo
  {journal} {Appl. Phys. Lett.}\ }\textbf {\bibinfo {volume} {116}},\ \bibinfo
  {pages} {013701} (\bibinfo {year} {2020})}\BibitemShut {NoStop}%
\bibitem [{\citenamefont {Tijore}\ \emph {et~al.}(2020)\citenamefont {Tijore},
  \citenamefont {Margadant}, \citenamefont {Yao}, \citenamefont {Hariharan},
  \citenamefont {Chew}, \citenamefont {Powell}, \citenamefont {Bonney},\ and\
  \citenamefont {Sheetz}}]{tijore2020ultrasoundmediated}%
  \BibitemOpen
  \bibfield  {author} {\bibinfo {author} {\bibfnamefont {A.}~\bibnamefont
  {Tijore}}, \bibinfo {author} {\bibfnamefont {F.}~\bibnamefont {Margadant}},
  \bibinfo {author} {\bibfnamefont {M.}~\bibnamefont {Yao}}, \bibinfo {author}
  {\bibfnamefont {A.}~\bibnamefont {Hariharan}}, \bibinfo {author}
  {\bibfnamefont {C.~A.~Z.}\ \bibnamefont {Chew}}, \bibinfo {author}
  {\bibfnamefont {S.}~\bibnamefont {Powell}}, \bibinfo {author} {\bibfnamefont
  {G.~K.}\ \bibnamefont {Bonney}}, \ and\ \bibinfo {author} {\bibfnamefont
  {M.}~\bibnamefont {Sheetz}},\ }\bibfield  {title} {\enquote {\bibinfo {title}
  {Ultrasound-mediated mechanical forces selectively kill tumor cells},}\
  }\href {\doibase 10.1101/2020.10.09.332726} {\bibfield  {journal} {\bibinfo
  {journal} {bioRxiv}\ }\textbf {\bibinfo {volume} {10.1101}},\ \bibinfo
  {pages} {2020.10.09.332726} (\bibinfo {year} {2020})}\BibitemShut {NoStop}%
\bibitem [{\citenamefont {Miller}\ \emph {et~al.}(2005)\citenamefont {Miller},
  \citenamefont {Leor},\ and\ \citenamefont {Rubinsky}}]{miller2005cancer}%
  \BibitemOpen
  \bibfield  {author} {\bibinfo {author} {\bibfnamefont {L.}~\bibnamefont
  {Miller}}, \bibinfo {author} {\bibfnamefont {J.}~\bibnamefont {Leor}}, \ and\
  \bibinfo {author} {\bibfnamefont {B.}~\bibnamefont {Rubinsky}},\ }\bibfield
  {title} {\enquote {\bibinfo {title} {Cancer {{Cells Ablation}} with
  {{Irreversible Electroporation}}},}\ }\href {\doibase
  10.1177/153303460500400615} {\bibfield  {journal} {\bibinfo  {journal}
  {Technol. Cancer Res. Treat.}\ }\textbf {\bibinfo {volume} {4}},\ \bibinfo
  {pages} {699--705} (\bibinfo {year} {2005})}\BibitemShut {NoStop}%
\bibitem [{\citenamefont {Perkons}\ \emph {et~al.}(2018)\citenamefont
  {Perkons}, \citenamefont {Stein}, \citenamefont {Nwaezeapu}, \citenamefont
  {Wildenberg}, \citenamefont {Saleh}, \citenamefont {{Itkin-Ofer}},
  \citenamefont {Ackerman}, \citenamefont {Soulen}, \citenamefont {Hunt},
  \citenamefont {Nadolski},\ and\ \citenamefont
  {Gade}}]{perkons2018electrolytic}%
  \BibitemOpen
  \bibfield  {author} {\bibinfo {author} {\bibfnamefont {N.~R.}\ \bibnamefont
  {Perkons}}, \bibinfo {author} {\bibfnamefont {E.~J.}\ \bibnamefont {Stein}},
  \bibinfo {author} {\bibfnamefont {C.}~\bibnamefont {Nwaezeapu}}, \bibinfo
  {author} {\bibfnamefont {J.~C.}\ \bibnamefont {Wildenberg}}, \bibinfo
  {author} {\bibfnamefont {K.}~\bibnamefont {Saleh}}, \bibinfo {author}
  {\bibfnamefont {R.}~\bibnamefont {{Itkin-Ofer}}}, \bibinfo {author}
  {\bibfnamefont {D.}~\bibnamefont {Ackerman}}, \bibinfo {author}
  {\bibfnamefont {M.~C.}\ \bibnamefont {Soulen}}, \bibinfo {author}
  {\bibfnamefont {S.~J.}\ \bibnamefont {Hunt}}, \bibinfo {author}
  {\bibfnamefont {G.~J.}\ \bibnamefont {Nadolski}}, \ and\ \bibinfo {author}
  {\bibfnamefont {T.~P.}\ \bibnamefont {Gade}},\ }\bibfield  {title} {\enquote
  {\bibinfo {title} {Electrolytic ablation enables cancer cell targeting
  through {{pH}} modulation},}\ }\href {\doibase 10.1038/s42003-018-0047-1}
  {\bibfield  {journal} {\bibinfo  {journal} {Commun. Biol.}\ }\textbf
  {\bibinfo {volume} {1}},\ \bibinfo {pages} {48} (\bibinfo {year}
  {2018})}\BibitemShut {NoStop}%
\bibitem [{\citenamefont {Knavel}\ and\ \citenamefont
  {Brace}(2013)}]{knavel2013tumor}%
  \BibitemOpen
  \bibfield  {author} {\bibinfo {author} {\bibfnamefont {E.~M.}\ \bibnamefont
  {Knavel}}\ and\ \bibinfo {author} {\bibfnamefont {C.~L.}\ \bibnamefont
  {Brace}},\ }\bibfield  {title} {\enquote {\bibinfo {title} {Tumor
  {{Ablation}}: {{Common Modalities}} and {{General Practices}}},}\ }\href
  {\doibase 10.1053/j.tvir.2013.08.002} {\bibfield  {journal} {\bibinfo
  {journal} {Tech. Vasc. Interv. Radiol.}\ }\textbf {\bibinfo {volume} {16}},\
  \bibinfo {pages} {192--200} (\bibinfo {year} {2013})}\BibitemShut {NoStop}%
\bibitem [{\citenamefont {Ranft}\ \emph {et~al.}(2012)\citenamefont {Ranft},
  \citenamefont {Prost}, \citenamefont {J{\"u}licher},\ and\ \citenamefont
  {Joanny}}]{ranft2012tissue}%
  \BibitemOpen
  \bibfield  {author} {\bibinfo {author} {\bibfnamefont {J.}~\bibnamefont
  {Ranft}}, \bibinfo {author} {\bibfnamefont {J.}~\bibnamefont {Prost}},
  \bibinfo {author} {\bibfnamefont {F.}~\bibnamefont {J{\"u}licher}}, \ and\
  \bibinfo {author} {\bibfnamefont {J.-F.}\ \bibnamefont {Joanny}},\ }\bibfield
   {title} {\enquote {\bibinfo {title} {Tissue dynamics with permeation},}\
  }\href {\doibase 10.1140/epje/i2012-12046-5} {\bibfield  {journal} {\bibinfo
  {journal} {Eur. Phys. J. E}\ }\textbf {\bibinfo {volume} {35}},\ \bibinfo
  {pages} {46} (\bibinfo {year} {2012})}\BibitemShut {NoStop}%
\bibitem [{\citenamefont {Dolega}\ \emph {et~al.}(2017)\citenamefont {Dolega},
  \citenamefont {Delarue}, \citenamefont {Ingremeau}, \citenamefont {Prost},
  \citenamefont {Delon},\ and\ \citenamefont {Cappello}}]{dolega2017celllike}%
  \BibitemOpen
  \bibfield  {author} {\bibinfo {author} {\bibfnamefont {M.~E.}\ \bibnamefont
  {Dolega}}, \bibinfo {author} {\bibfnamefont {M.}~\bibnamefont {Delarue}},
  \bibinfo {author} {\bibfnamefont {F.}~\bibnamefont {Ingremeau}}, \bibinfo
  {author} {\bibfnamefont {J.}~\bibnamefont {Prost}}, \bibinfo {author}
  {\bibfnamefont {A.}~\bibnamefont {Delon}}, \ and\ \bibinfo {author}
  {\bibfnamefont {G.}~\bibnamefont {Cappello}},\ }\bibfield  {title} {\enquote
  {\bibinfo {title} {Cell-like pressure sensors reveal increase of mechanical
  stress towards the core of multicellular spheroids under compression},}\
  }\href {\doibase 10.1038/ncomms14056} {\bibfield  {journal} {\bibinfo
  {journal} {Nat. Commun.}\ }\textbf {\bibinfo {volume} {8}},\ \bibinfo {pages}
  {14056} (\bibinfo {year} {2017})}\BibitemShut {NoStop}%
\bibitem [{\citenamefont {Basan}\ \emph {et~al.}(2009)\citenamefont {Basan},
  \citenamefont {Risler}, \citenamefont {Joanny}, \citenamefont
  {Sastre-Garau},\ and\ \citenamefont {Prost}}]{basan2009homeostatic}%
  \BibitemOpen
  \bibfield  {author} {\bibinfo {author} {\bibfnamefont {M.}~\bibnamefont
  {Basan}}, \bibinfo {author} {\bibfnamefont {T.}~\bibnamefont {Risler}},
  \bibinfo {author} {\bibfnamefont {J.-F.}\ \bibnamefont {Joanny}}, \bibinfo
  {author} {\bibfnamefont {X.}~\bibnamefont {Sastre-Garau}}, \ and\ \bibinfo
  {author} {\bibfnamefont {J.}~\bibnamefont {Prost}},\ }\bibfield  {title}
  {\enquote {\bibinfo {title} {Homeostatic competition drives tumor growth and
  metastasis nucleation},}\ }\href {\doibase 10.2976/1.3086732} {\bibfield
  {journal} {\bibinfo  {journal} {HFSP J.}\ }\textbf {\bibinfo {volume} {3}},\
  \bibinfo {pages} {265--272} (\bibinfo {year} {2009})}\BibitemShut {NoStop}%
\bibitem [{\citenamefont {Simha}\ and\ \citenamefont
  {Ramaswamy}(2002)}]{simha2002hydrodynamic}%
  \BibitemOpen
  \bibfield  {author} {\bibinfo {author} {\bibfnamefont {R.}~\bibnamefont
  {Simha}}\ and\ \bibinfo {author} {\bibfnamefont {S.}~\bibnamefont
  {Ramaswamy}},\ }\bibfield  {title} {\enquote {\bibinfo {title} {Hydrodynamic
  {{Fluctuations}} and {{Instabilities}} in {{Ordered Suspensions}} of
  {{Self}}-{{Propelled Particles}}},}\ }\href {\doibase
  10.1103/PhysRevLett.89.058101} {\bibfield  {journal} {\bibinfo  {journal}
  {Phys. Rev. Lett.}\ }\textbf {\bibinfo {volume} {89}},\ \bibinfo {pages}
  {058101} (\bibinfo {year} {2002})}\BibitemShut {NoStop}%
\bibitem [{\citenamefont {Bittig}\ \emph {et~al.}(2008)\citenamefont {Bittig},
  \citenamefont {Wartlick}, \citenamefont {Kicheva}, \citenamefont
  {{Gonz{\'a}lez-Gait{\'a}n}},\ and\ \citenamefont
  {J{\"u}licher}}]{bittig2008dynamics}%
  \BibitemOpen
  \bibfield  {author} {\bibinfo {author} {\bibfnamefont {T.}~\bibnamefont
  {Bittig}}, \bibinfo {author} {\bibfnamefont {O.}~\bibnamefont {Wartlick}},
  \bibinfo {author} {\bibfnamefont {A.}~\bibnamefont {Kicheva}}, \bibinfo
  {author} {\bibfnamefont {M.}~\bibnamefont {{Gonz{\'a}lez-Gait{\'a}n}}}, \
  and\ \bibinfo {author} {\bibfnamefont {F.}~\bibnamefont {J{\"u}licher}},\
  }\bibfield  {title} {\enquote {\bibinfo {title} {Dynamics of anisotropic
  tissue growth},}\ }\href {\doibase 10.1088/1367-2630/10/6/063001} {\bibfield
  {journal} {\bibinfo  {journal} {New J. Phys.}\ }\textbf {\bibinfo {volume}
  {10}},\ \bibinfo {pages} {063001} (\bibinfo {year} {2008})}\BibitemShut
  {NoStop}%
\bibitem [{\citenamefont {Blackiston}\ \emph {et~al.}(2009)\citenamefont
  {Blackiston}, \citenamefont {McLaughlin},\ and\ \citenamefont
  {Levin}}]{blackiston2009bioelectric}%
  \BibitemOpen
  \bibfield  {author} {\bibinfo {author} {\bibfnamefont {D.~J.}\ \bibnamefont
  {Blackiston}}, \bibinfo {author} {\bibfnamefont {K.~A.}\ \bibnamefont
  {McLaughlin}}, \ and\ \bibinfo {author} {\bibfnamefont {M.}~\bibnamefont
  {Levin}},\ }\bibfield  {title} {\enquote {\bibinfo {title} {Bioelectric
  controls of cell proliferation: {{Ion}} channels, membrane voltage and the
  cell cycle},}\ }\href {\doibase 10.4161/cc.8.21.9888} {\bibfield  {journal}
  {\bibinfo  {journal} {Cell Cycle}\ }\textbf {\bibinfo {volume} {8}},\
  \bibinfo {pages} {3527--3536} (\bibinfo {year} {2009})}\BibitemShut {NoStop}%
\bibitem [{\citenamefont {Darcy}(1856)}]{darcy1856fontaines}%
  \BibitemOpen
  \bibfield  {author} {\bibinfo {author} {\bibfnamefont {H.~P.~G.}\
  \bibnamefont {Darcy}},\ }\href@noop {} {\emph {\bibinfo {title} {Les
  Fontaines Publiques de La Ville de {{Dijon}}}}}\ (\bibinfo  {publisher}
  {{Dalmont}},\ \bibinfo {address} {{Paris, France}},\ \bibinfo {year}
  {1856})\BibitemShut {NoStop}%
\bibitem [{\citenamefont {Ramaswamy}\ \emph {et~al.}(2000)\citenamefont
  {Ramaswamy}, \citenamefont {Toner},\ and\ \citenamefont
  {Prost}}]{ramaswamy2000nonequilibrium}%
  \BibitemOpen
  \bibfield  {author} {\bibinfo {author} {\bibfnamefont {S.}~\bibnamefont
  {Ramaswamy}}, \bibinfo {author} {\bibfnamefont {J.}~\bibnamefont {Toner}}, \
  and\ \bibinfo {author} {\bibfnamefont {J.}~\bibnamefont {Prost}},\ }\bibfield
   {title} {\enquote {\bibinfo {title} {Nonequilibrium {{Fluctuations}},
  {{Traveling Waves}}, and {{Instabilities}} in {{Active Membranes}}},}\ }\href
  {\doibase 10.1103/PhysRevLett.84.3494} {\bibfield  {journal} {\bibinfo
  {journal} {Phys. Rev. Lett.}\ }\textbf {\bibinfo {volume} {84}},\ \bibinfo
  {pages} {3494--3497} (\bibinfo {year} {2000})}\BibitemShut {NoStop}%
\bibitem [{\citenamefont {Kirby}(2013)}]{kirby2013micro}%
  \BibitemOpen
  \bibfield  {author} {\bibinfo {author} {\bibfnamefont {B.~J.}\ \bibnamefont
  {Kirby}},\ }\href@noop {} {\emph {\bibinfo {title} {Micro- and {{Nanoscale
  Fluid Mechanics}}}}}\ (\bibinfo  {publisher} {{Cambridge University Press}},\
  \bibinfo {address} {{Cambridge, England}},\ \bibinfo {year}
  {2013})\BibitemShut {NoStop}%
\bibitem [{\citenamefont {Delarue}\ \emph {et~al.}(2014)\citenamefont
  {Delarue}, \citenamefont {Joanny}, \citenamefont {J{\"u}licher},\ and\
  \citenamefont {Prost}}]{delarue2014stress}%
  \BibitemOpen
  \bibfield  {author} {\bibinfo {author} {\bibfnamefont {M.}~\bibnamefont
  {Delarue}}, \bibinfo {author} {\bibfnamefont {J.-F.}\ \bibnamefont {Joanny}},
  \bibinfo {author} {\bibfnamefont {F.}~\bibnamefont {J{\"u}licher}}, \ and\
  \bibinfo {author} {\bibfnamefont {J.}~\bibnamefont {Prost}},\ }\bibfield
  {title} {\enquote {\bibinfo {title} {Stress distributions and cell flows in a
  growing cell aggregate},}\ }\href {\doibase 10.1098/rsfs.2014.0033}
  {\bibfield  {journal} {\bibinfo  {journal} {Interface Focus}\ }\textbf
  {\bibinfo {volume} {4}},\ \bibinfo {pages} {20140033} (\bibinfo {year}
  {2014})}\BibitemShut {NoStop}%
\bibitem [{\citenamefont {Forgacs}\ \emph {et~al.}(1998)\citenamefont
  {Forgacs}, \citenamefont {Foty}, \citenamefont {Shafrir},\ and\ \citenamefont
  {Steinberg}}]{forgacs1998viscoelastic}%
  \BibitemOpen
  \bibfield  {author} {\bibinfo {author} {\bibfnamefont {G.}~\bibnamefont
  {Forgacs}}, \bibinfo {author} {\bibfnamefont {R.~A.}\ \bibnamefont {Foty}},
  \bibinfo {author} {\bibfnamefont {Y.}~\bibnamefont {Shafrir}}, \ and\
  \bibinfo {author} {\bibfnamefont {M.~S.}\ \bibnamefont {Steinberg}},\
  }\bibfield  {title} {\enquote {\bibinfo {title} {Viscoelastic {{Properties}}
  of {{Living Embryonic Tissues}}: A {{Quantitative Study}}},}\ }\href
  {\doibase 10.1016/S0006-3495(98)77932-9} {\bibfield  {journal} {\bibinfo
  {journal} {Biophys. J.}\ }\textbf {\bibinfo {volume} {74}},\ \bibinfo {pages}
  {2227--2234} (\bibinfo {year} {1998})}\BibitemShut {NoStop}%
\bibitem [{\citenamefont {Netti}\ \emph {et~al.}(2000)\citenamefont {Netti},
  \citenamefont {Berk}, \citenamefont {Swartz}, \citenamefont {Grodzinsky},\
  and\ \citenamefont {Jain}}]{netti2000role}%
  \BibitemOpen
  \bibfield  {author} {\bibinfo {author} {\bibfnamefont {P.~A.}\ \bibnamefont
  {Netti}}, \bibinfo {author} {\bibfnamefont {D.~A.}\ \bibnamefont {Berk}},
  \bibinfo {author} {\bibfnamefont {M.~A.}\ \bibnamefont {Swartz}}, \bibinfo
  {author} {\bibfnamefont {A.~J.}\ \bibnamefont {Grodzinsky}}, \ and\ \bibinfo
  {author} {\bibfnamefont {R.~K.}\ \bibnamefont {Jain}},\ }\bibfield  {title}
  {\enquote {\bibinfo {title} {Role of {{Extracellular Matrix Assembly}} in
  {{Interstitial Transport}} in {{Solid Tumors}}},}\ }\href {\doibase
  http://cancerres.aacrjournals.org/content/60/9/2497} {\bibfield  {journal}
  {\bibinfo  {journal} {Cancer Res.}\ }\textbf {\bibinfo {volume} {60}},\
  \bibinfo {pages} {2497} (\bibinfo {year} {2000})}\BibitemShut {NoStop}%
\bibitem [{\citenamefont {Brace}\ and\ \citenamefont
  {Guyton}(1977)}]{brace1977interaction}%
  \BibitemOpen
  \bibfield  {author} {\bibinfo {author} {\bibfnamefont {R.~A.}\ \bibnamefont
  {Brace}}\ and\ \bibinfo {author} {\bibfnamefont {A.~C.}\ \bibnamefont
  {Guyton}},\ }\bibfield  {title} {\enquote {\bibinfo {title} {Interaction of
  transcapillary {{Starling}} forces in the isolated dog forelimb},}\ }\href
  {\doibase 10.1152/ajpheart.1977.233.1.H136} {\bibfield  {journal} {\bibinfo
  {journal} {Am. J. Physiol.}\ }\textbf {\bibinfo {volume} {233}},\ \bibinfo
  {pages} {H136--H140} (\bibinfo {year} {1977})}\BibitemShut {NoStop}%
\bibitem [{\citenamefont {Erez}\ and\ \citenamefont
  {Shitzer}(1980)}]{erez1980controlled}%
  \BibitemOpen
  \bibfield  {author} {\bibinfo {author} {\bibfnamefont {A.}~\bibnamefont
  {Erez}}\ and\ \bibinfo {author} {\bibfnamefont {A.}~\bibnamefont {Shitzer}},\
  }\bibfield  {title} {\enquote {\bibinfo {title} {Controlled {{Destruction}}
  and {{Temperature Distributions}} in {{Biological Tissues Subjected}} to
  {{Monoactive Electrocoagulation}}},}\ }\href {\doibase 10.1115/1.3138197}
  {\bibfield  {journal} {\bibinfo  {journal} {J. Biomech. Eng.}\ }\textbf
  {\bibinfo {volume} {102}},\ \bibinfo {pages} {42--49} (\bibinfo {year}
  {1980})}\BibitemShut {NoStop}%
\bibitem [{\citenamefont {Hirschhaeuser}\ \emph {et~al.}(2010)\citenamefont
  {Hirschhaeuser}, \citenamefont {Menne}, \citenamefont {Dittfeld},
  \citenamefont {West}, \citenamefont {{Mueller-Klieser}},\ and\ \citenamefont
  {{Kunz-Schughart}}}]{hirschhaeuser2010multicellular}%
  \BibitemOpen
  \bibfield  {author} {\bibinfo {author} {\bibfnamefont {F.}~\bibnamefont
  {Hirschhaeuser}}, \bibinfo {author} {\bibfnamefont {H.}~\bibnamefont
  {Menne}}, \bibinfo {author} {\bibfnamefont {C.}~\bibnamefont {Dittfeld}},
  \bibinfo {author} {\bibfnamefont {J.}~\bibnamefont {West}}, \bibinfo {author}
  {\bibfnamefont {W.}~\bibnamefont {{Mueller-Klieser}}}, \ and\ \bibinfo
  {author} {\bibfnamefont {L.~A.}\ \bibnamefont {{Kunz-Schughart}}},\
  }\bibfield  {title} {\enquote {\bibinfo {title} {Multicellular tumor
  spheroids: {{An}} underestimated tool is catching up again},}\ }\href
  {\doibase 10.1016/j.jbiotec.2010.01.012} {\bibfield  {journal} {\bibinfo
  {journal} {J. Biotechnol.}\ }\textbf {\bibinfo {volume} {148}},\ \bibinfo
  {pages} {3--15} (\bibinfo {year} {2010})}\BibitemShut {NoStop}%
\bibitem [{\citenamefont {Wang}\ \emph {et~al.}(2017)\citenamefont {Wang},
  \citenamefont {Stone},\ and\ \citenamefont {Golestanian}}]{wang2017shape}%
  \BibitemOpen
  \bibfield  {author} {\bibinfo {author} {\bibfnamefont {X.}~\bibnamefont
  {Wang}}, \bibinfo {author} {\bibfnamefont {H.~A.}\ \bibnamefont {Stone}}, \
  and\ \bibinfo {author} {\bibfnamefont {R.}~\bibnamefont {Golestanian}},\
  }\bibfield  {title} {\enquote {\bibinfo {title} {Shape of the growing front
  of biofilms},}\ }\href {\doibase 10.1088/1367-2630/aa983f} {\bibfield
  {journal} {\bibinfo  {journal} {New J. Phys.}\ }\textbf {\bibinfo {volume}
  {19}},\ \bibinfo {pages} {125007} (\bibinfo {year} {2017})}\BibitemShut
  {NoStop}%
\bibitem [{\citenamefont {Martin}\ and\ \citenamefont
  {Risler}(2021)}]{martin2021viscocapillary}%
  \BibitemOpen
  \bibfield  {author} {\bibinfo {author} {\bibfnamefont {M.}~\bibnamefont
  {Martin}}\ and\ \bibinfo {author} {\bibfnamefont {T.}~\bibnamefont
  {Risler}},\ }\bibfield  {title} {\enquote {\bibinfo {title} {Viscocapillary
  instability in cellular spheroids},}\ }\href {\doibase
  10.1088/1367-2630/abe9d6} {\bibfield  {journal} {\bibinfo  {journal} {New J.
  Phys.}\ }\textbf {\bibinfo {volume} {23}},\ \bibinfo {pages} {033032}
  (\bibinfo {year} {2021})}\BibitemShut {NoStop}%
\bibitem [{\citenamefont {Hanahan}\ and\ \citenamefont
  {Weinberg}(2011)}]{hanahan2011hallmarks}%
  \BibitemOpen
  \bibfield  {author} {\bibinfo {author} {\bibfnamefont {D.}~\bibnamefont
  {Hanahan}}\ and\ \bibinfo {author} {\bibfnamefont {R.~A.}\ \bibnamefont
  {Weinberg}},\ }\bibfield  {title} {\enquote {\bibinfo {title} {Hallmarks of
  {{Cancer}}: {{The Next Generation}}},}\ }\href {\doibase
  10.1016/j.cell.2011.02.013} {\bibfield  {journal} {\bibinfo  {journal}
  {Cell}\ }\textbf {\bibinfo {volume} {144}},\ \bibinfo {pages} {646--674}
  (\bibinfo {year} {2011})}\BibitemShut {NoStop}%
\end{thebibliography}%

\end{document}